\documentclass[apjl,twocolumn,floats]{emulateapj}
\pdfoutput=1
\usepackage{bm}
\usepackage{multirow}
\usepackage{color}
\usepackage{hyperref}
\bibpunct[; ]{(}{)}{;}{a}{}{;}

\usepackage{amsmath}
\usepackage{amsfonts}
\usepackage{amssymb}
\usepackage{epsfig}
\usepackage{hyperref}
\usepackage{graphicx}

\newcommand{\be}{\begin{equation}}
\newcommand{\ee}{\end{equation}}
\newcommand{\ba}{\begin{eqnarray}}
\newcommand{\ea}{\end{eqnarray}}

\begin{document}

\title{Prospects for detecting CII emission during the Epoch of Reionization}

\author{Marta Silva$^{1,2,3,5}$,Mario G. Santos$^{3,4,1}$, Asantha Cooray$^2$,Yan Gong$^{6,2}$}
\affil{$^1$CENTRA, Instituto Superior T\'ecnico, Technical University of Lisbon, Lisboa 1049-001, Portugal}
\affil{$^2$Department of Physics \& Astronomy, University of California, Irvine, CA 92697}
\affil{$^3$Department of Physics \& Astronomy, University of Western Cape, Cape Town 753sistemáticos5, South Africa}
\affil{$^4$SKA SA, The Park, Park Road, Cape Town 7405, South Africa}
\affil{$^5$Kapteyn Astronomical Institute, The University of Groningen, Landleven 12, 9747 AD Groningen, The Netherland}
\affil{$^6$National Astronomical Observatories, Chinese Academy of Sciences, 20A Datun Road, Chaoyang District, Beijing 100012, China}
\begin{abstract}

We produce simulations of the atomic CII line emission in large sky fields in order to determine the current and future
prospects for mapping this line during the high redshift epoch of reionization. 
We calculate the CII line intensity, redshift evolution and spatial fluctuations using observational relations 
between CII emission and the galaxy star formation rate (SFR) over the frequency range 200 - 300 GHz. We estimate 
an averaged intensity of ${\rm I_{\rm CII}=(4 \pm 2)\times10^{2}\, Jy\, \rm sr^{-1}}$ in the redshift
range $z\, \sim\, 5.3\, -\, 8.5$. Observations of the CII emission in this frequency range will suffer 
contamination from emission lines at lower redshifts, in particular CO rotational lines. Using simulations, we 
estimated the CO contamination to be ${\rm I_{\rm CO} \approx 10^{3}\, Jy \,  sr^{-1}}$ (originating from galaxies 
at $z\, <\, 2.5$). Using detailed simulations 
of the CII and CO emission across a range of redshifts, we generate maps as a function of angle and 
frequency, fully taking into account this resolution and light cone effects. In order to reduce the foreground 
contamination we find that we should mask galaxies below redshifts $\sim 2.5$ with a CO(J:2-1) 
to CO(J:6-5) line flux density higher than ${\rm 5\times 10^{-22}\, W\ m^{-2}}$ or a AB magnitude lower than ${\rm m_{\rm K}\, =\, 22}$.  
We estimate that the additional continuum contamination originating in emission from stars and in dust, free-free, 
free-bound and two photon emission in the ISM is of the order of ${\rm 10^{5}\, Jy\, sr^{-1}}$ however it can be removed from observation 
due to the smooth evolution of this foreground with frequency. We 
also consider the possibility of cross correlating foreground lines with galaxy surveys in order to 
probe the intensity of the foregrounds. Finally, we discuss the expected constraints from two experiments capable 
of measuring the expected CII power spectrum.

\end{abstract}


\maketitle

\section{Introduction}

The epoch of reionization (EoR) is a fundamental stage in the history of large scale structure formation. 
The process of hydrogen ionization was fueled by radiation from the first galaxies which formed in overdense regions. Therefore, this process depends on a large set of astrophysical and cosmological parameters 
\citep{2000ApJ...537...55V}.

There are already several experiments in operation using low frequency telescopes, such as the Murchison Widefield Arrays (MWA) \citep{2013PASA...30....7T}, the Giant Metrewave Radio Telescope (GMRT) \citep{2011MNRAS.413.1174P}, the Precision Array for Probing the Epoch of Reionization (PAPER) \citep{2010AJ....139.1468P} and the Low Frequency Array (LOFAR) \citep{2006astro.ph.10596R} aimed at constraining this epoch through the measurement of the 21cm signal. Future experiments such as Phase II of HERA \footnote{http://reionization.org} and the Square Kilometre Array low frequency instrument (SKA1-LOW) \citep{2013ExA....36..235M} should push this measurement to even higher redshifts.

One of the main challenges for probing the EoR with the 21 cm line is that observations will be contaminated by foregrounds several orders of magnitude higher than the signal \citep{1999A&A...345..380S}.  Although the frequency smoothness of these foregrounds provides a way to remove them \citep{2005ApJ...625..575S, 2012MNRAS.423.2518C}, the combination with calibration errors and systematics complicates the foreground cleaning process.

Independent ways to measure this signal and to probe the reionization process are therefore required in order to ensure the validity of our measurements related to reionization. 

In this work we analyse the use of  CII intensity mapping both to probe the EoR during its final stages and to confirm and complement the 21 cm data. Although not resolving individual sources, the intensity mapping technique has the advantage of measuring all the 
emission in a given frequency band originating from a relatively large sky patch. This way, it is sensitive to radiation from faint sources and the diffuse IGM which at these high redshifts cannot be detected with other methods, but whose contribution to the total signal is often important \citep{2012ApJ...745...49G,2012arXiv1205.1493S}.  Compared to other techniques, intensity mapping has the advantage of providing three dimensional spatial information of the sources of emission that can be used to further understand the processes of structure formation. Intensity maps can also be used as cosmological probes since the fluctuations in the intensity of emission/absorption lines are correlated with the underlying dark matter density fluctuations \citep{2011ApJ...730L..30C}. In particular, with CII, we can make maps of the sources of ionization, while the 21cm signal will simply be sensitive to the IGM.

We show the potential of CII intensity mapping by simulating mock observational cones of CII emission 
and its foregrounds at frequencies from 200 GHz to 300 GHz, taking into account the light cone effects. This allows us to test possible ways to reduce the foregrounds without erasing the signal. The main foregrounds in CII intensity maps from the EoR will be contamination from other far-infrared emission lines from lower redshifts, in particular emission from CO rotational transitions. CO emission from normal galaxies at ${\rm z > 1}$ is poorly constrained by observations. Therefore, in order to properly estimate the intensity of these lines and the contamination power spectra relative to CII observations, we used two independent methods. First we used the simulated galaxies catalog from the SAX-Sky simulation which uses a phenomenological model to calculate the luminosities of different CO transitions \citep{2009ApJ...702.1321O}. We then confirmed our predictions using IR luminosity functions (LFs) and other observational data to estimate the relative 
intensities of the several CO transitions.  

We find that CO contamination is dominated by bright sources and so it can be efficiently reduced by masking the pixels where radiation from these sources is observed.  In order to apply this procedure we need a complementary experiment to measure CO emission from galaxies brighter than a given flux. This can be done with galaxy surveys targeting the CO emission, which would on its own be a powerful astrophysical probe on the conditions of the ISM. Alternatively the masking of the contaminant galaxies could be done with a CO tracer, easier to be observed, such as the SFR or the relative magnitude in a given filter.
 
We also explore 
the possibility of cross correlating CII and 21 cm maps in order to obtain maps of the EoR that are clean from foregrounds and systematics. This is possible since two lines 
emitted from the same redshift will be observed at different frequencies and so they will be contaminated mainly by uncorrelated foregrounds \citep{2012ApJ...745...49G}. 

This paper is organized as follows: In Section 2 we describe how to theoretically estimate the CII emission. In 
Section 3 we describe the CII foregrounds. In Sections 4 and 5 we describe how we used simulations to generate the signal and the foregrounds. In the 5th Section we present the parameters of an experiment able to measure the CII signal and the CO signal in the 200 GHz to 300 GHz band and in Section 6 we discuss how to remove the CII foregrounds. We conclude with a discussion of the results obtained in Section 7.

\section{Calculating CII emission} 

CII emission is originated in i) the interstellar medium (ISM), ii) Photodissociation regions (PDRs), iii) ionized regions (HII regions), iv) cold atomic gas or v) CO-dark molecular gas (regions in the boundary of molecular clouds with ${\rm H}_2$ but without CO gas). Observations of the relative intensity of different emission lines have shown that the main source of CII emission is the dense PDRs located in the boundary of HII regions. PDRs are dense and warm regions of the ISM located between HII regions and molecular clouds. They contain mostly neutral gas, but due to their proximity from O, B stars or AGNs the physical and chemical properties of the gas are set by the strong far ultraviolet (FUV) field. The strong FUV to X-ray radiation that penetrates the PDR is absorbed by dust grains which emit electrons heating the gas, or by atoms with an energy threshold for ionization below the Lyman alpha limit such as carbon, oxygen and nitrogen. The FUV also causes 
transitions from atomic to molecular hydrogen and from ionized carbon to carbon monoxide \citep{1997ARA&A..35..179H}. 

CII photons are emitted in PDRs as a cooling mechanism and so they are a consequence of pre-existing heat. CO-dark clouds are envelopes of dense ${\rm H_2}$ gas with densities too low for carbon to be converted to CO, but which can be identified by their CII emission. The contribution from CO-dark clouds to the total CII budget is not yet clear but recent studies of our galaxy indicate that it can be high (up to $\sim 28 \%$) \citep{2013A&A...554A.103P} under certain astrophysical conditions, more characteristic of the low redshift universe. Diffuse cold atomic gas can be characterized by its emission in the hydrogen 21 cm line and in the CII line. The intensity of emission in this gas phase will be proportional to the collisional rate which depends on the gas density and temperature and therefore also on FUV strength.

The carbon ionization energy is only 11.3 eV which is less than the 13.6 eV necessary to ionize hydrogen so at first we could expect, as was done in \citet{2012ApJ...745...49G}, that all the carbon in HII regions would be ionized. There would then be a high emission in the CII 157$\mu$m line since its excitation potential is only of 91 ${\rm K}$. Under this assumption most of the CII emission would come from the highest density locations inside HII regions. However this is not supported by observations: several observational maps of the spatial distribution of CII emission in galaxies indicate that the CII emission is mainly originating in PDRs and that HII regions contribute only a few $\%$ of the total CII emission \citep{2012A&A...548A..91L,2014ApJ...781L..15R}. There are studies which indicate a contribution from 
HII regions that can reach up to 30$\%$ of the total CII emission \citep{1994ApJ...423..223C,1999ESASP.427..973S,2003A&A...406..155A,2013MNRAS.434.2051R}. However, these studies point out that most CII emission is originated in the low density HII regions. The more simple explanation for this unexpected result is that the carbon in the more dense places in HII regions is highly shielded from radiation by hydrogen and so almost all of the ionized carbon is located in low density regions.

The \textit{Herschel} telescope and the SOFIA observatory were used to observe typical tracers of PDRs, HII regions and 
other galactic regions \citep{2013A&A...556A..92K} and these observations showed that CII emission has a more 
complex spatial structure than most other infrared lines. In order to properly estimate the CII 
emission from a galaxy, it is necessary to observe it with high spatial resolution, which is not possible for most distant galaxies. Alternatively we can use the intensity mapping technique to measure the integrated CII emission from many galaxies.   

\subsection{Theoretical formulas to estimate CII emission}
\label{subsec: LCII_versus_SFR}
 
The intensity of CII emission is given theoretically \citep{2012ApJ...745...49G} as:
\ba
\label{eq:ICII_teorico}
I_{\nu} = \frac{hc}{4\pi H(z)(1+z)^3}A_{\rm ul}\ f_{\rm CII}^{\rm grd}n_{\rm CII}(z)\times \\ \nonumber
\frac{g_{\rm u}}{g_{\rm l}}{\rm exp}(-T_{\star,\rm ul}/T_{S,\rm ul})\bigg[ 1-\frac{{\rm exp}(T_{\star,\rm ul}/T_{S,\rm ul})-1}{(2h\nu^3/c^2I_{\nu})_{\nu_{\rm ul}}} \bigg],
\ea
\\
where $f_{\rm CII}^{\rm grd}$ is the fraction of CII ions at the ground level $^2P_{1/2}$, $n_{\rm CII}$ is the number 
density of once ionized carbon atoms, H(z) is the hubble parameter, $T_{S}$ 
is the spin temperature and $T_{\star}\equiv h\nu_{\rm ul}/k_B$ (where $\nu_{\rm ul}$ is the frequency of the transition).
The statistical weights are $g_{\rm u}=4$ and $g_{\rm l}=2$ and the Einstein spontaneous emission coefficient 
is $A_{\rm ul}=2.36\times10^{-6} s^{-1}$.

As many of the parameters in Equation \ref{eq:ICII_teorico} are poorly known and cannot be directly obtained from observations we use an alternative formula, based on the halo model, to obtain the intensity of a line emitted from several galaxies in a relatively large volume. For this we made the simplification of assuming that the average luminosity of each of these galaxies is only a function of the mass of the dark matter 
halo which contains it and at most its redshift. The average intensity of a line is then given by:
\be
\label{eq:Int_CII}
\bar{I}(z)=\int^{\rm M_{\rm max}} _{\rm M_{\rm min}}dM\frac{dn}{dM}\frac{L(M,z)}{4\pi D_{\rm L}^{\rm 2}}y(z)D^2_A
\ee
\\
where $dn/dM$ is the halo mass function \citep{1999MNRAS.308..119S}, $M$ is the halo mass, 
$M_{\rm min}\, =\, 10^{\rm 8}\ {\rm M_{\odot}}$, $M_{\rm max}\, =\, 10^{14}\ {\rm M_{\odot}}$, $D_{\rm L}$ is the proper luminosity distance,
$D_{\rm A}$ is the comoving angular diameter distance and $y(z)=d\chi/d\nu$, where $\chi$ is the comoving distance 
and $\nu$ is the observed frequency.
The relation between $L_{\rm CII}$ and the halo mass is physically based in the dependence of $L_{\rm CII}$ in the 
number density of CII atoms which should be proportional to the halo mass.

\subsection{Calculating CII emission using observational based relations}
\label{subsec: LCII observational}

The CII luminosity of a galaxy can be estimated from other observable quantities as long as there is a reasonable
correlation between the two. For large volumes, since we integrate over several galaxies, it is even more reliable to use these observational relations to estimate the overall luminosity from these regions.
CII emission is powered by FUV radiation and so there is a correlation between these two quantities which can be 
converted to a relation between CII and FIR luminosities given that in the star forming galaxies (which will dominate the signal) there is a known correlation 
between the FUV and FIR fluxes.
The CII luminosity of a galaxy also depends on other astrophysical properties of the galaxy such as its 
metallicity, however the average ratio $R=\frac{L_{\rm CII}}{L_{\rm FIR}}$ for nearby, 
late type galaxies and for $10^8\, {\rm L}_{\odot} \le L_{\rm FIR} \le 10^{10.5}\, {\rm L}_{\odot}$ is approximately 
constant \citep{2002A&A...385..454B} and is given by:
\be
L_{\rm CII(M,z)}\left[{\rm L_{\odot}} \right]=0.003 \times L_{\rm FIR}.
\label{eq:LCII_LFIR}
\ee

This relation is also consistent with recent observations of high redshift galaxies \citep{2010ApJ...724..957S} 
and with observations of ULIRGS ($L_{\rm IR}>10^{11.5}\, {\rm L}_{\odot}$), 
where a ratio of $R=(0.8-4)\times10^{-3}$ in the CII to FIR luminosities was found \citep{2014ApJ...781L..15R}.
In PDRs the same ratio is inversely proportional to the strength of the 
ambient radiation field $G_0$, since $L_{\rm FIR}$ is proportional to $G_0$ and $L_{\rm CII}$ depends weakly on $G_0$ 
\citep{1999ApJ...527..795K}. Therefore, this ratio is likely to slightly increase to low mass galaxies 
(up to $R\sim 0.01$) and to decrease to high mass galaxies. 
The IR and the FIR luminosities are connected by the following relation: 
\be
L_{\rm IR}({\rm 8-1000\ \mu m}) = (1.89 \pm 0.26)L_{\rm FIR}({\rm 40-120\ \mu m}),
\label{eq:LIR_LFIR}
\ee
\\
obtained by \citet{2003ApJ...584...76C} using the IRAS Bright Galaxies Sample from \citet{1989AJ.....98..766S}.

The integrated IR 
luminosity, $L_{\rm IR} = L(8 - 1000 {\rm \mu m})$ is related to the galaxies star formation rate ($\psi$) by the \citet{1998ApJ...498..541K} relation:
\be
L_{\rm IR}(M,z) \left[ {\rm L_{\odot}} \right] =5.8 \times 10^{\rm 9} \psi(M,z) \left[{\rm M_{\odot}/yr} \right].
\label{eq:L_IR_SFR}
\ee
Using Equations \ref{eq:LCII_LFIR}, \ref{eq:L_IR_SFR} and \ref{eq:LIR_LFIR} we obtained the following relation 
between CII luminosity and SFR:
\ba                        
\label{eq:L_SFR}
L_{\rm CII(M,z)}[{\rm L_{\odot}}] &=& 0.003 \times L_{\rm FIR}\left[{\rm L_{\odot}}\right]\\ \nonumber
                                  &=& 0.003 \times 0.53 \times L_{\rm IR} \left[ {\rm L_{\odot} } \right]\\ \nonumber
                                  &=& 9.22 \times 10^6 \psi(M,z)\left[{\rm M_{\odot}/yr}\right].
\ea

The connection between CII luminosity and the SFR can be easily understood in the case of CII emission arising 
from warm photodissociating regions, since in this case the FUV radiation ionizes the carbon in the outer 
layers of the photon-dominated molecular clumps which, in its turn emits CII with a luminosity proportional to the
FUV flux which is linked to the galaxy SFR \citep{2011MNRAS.416.2712D}. In HII regions the amount of ionized carbon should 
increase with the size of the region, which is proportional to the stellar radiation UV intensity. However, given that not all carbon is necessarily ionized at the same time in HII regions and that the CII luminosity of these regions also depends on the astrophysical conditions of the gas, one expects a considerable dispersion in the relation between CII luminosity from HII regions and the SFR.
Alternative relations between the CII luminosity and the SFR, obtained using different 
galaxy datasets and using a SFR estimated from the infrared luminosity or from the H$\alpha$ luminosity, can be found for example in \citet{2002A&A...385..454B}, \citet{2011MNRAS.416.2712D} or \citet{2012ApJ...755..171S}. 
All of the referred observational studies indicate that the ratio between CII and SFR is smaller for ultra-luminous 
galaxies although these galaxies account for no more than a few percent of the total emission, which justifies our use of a constant ratio.

\begin{figure*}[!t]
\hspace{-10pt}
\includegraphics[scale = 0.46]{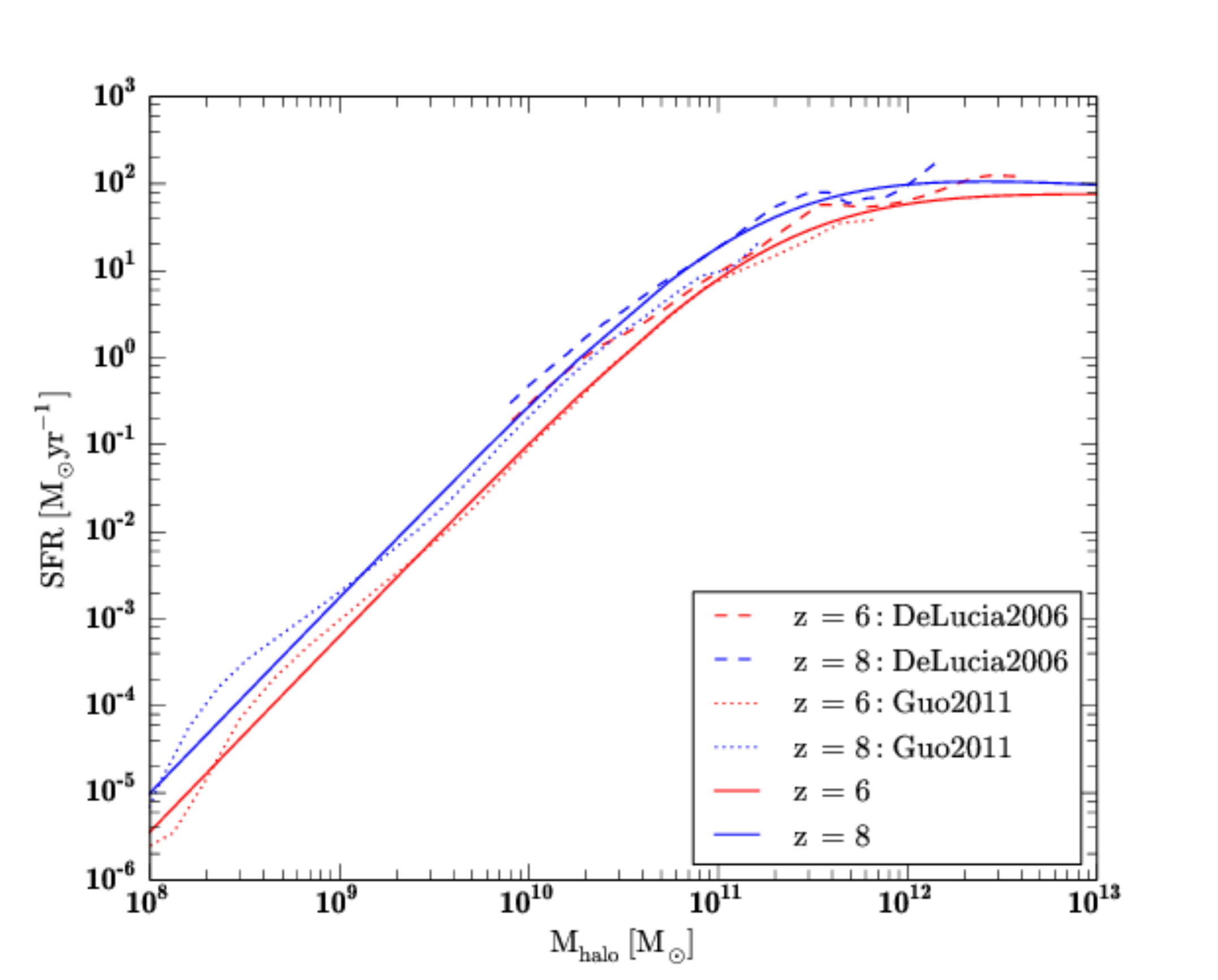} 
\hspace{-25pt}
\includegraphics[scale = 0.46]{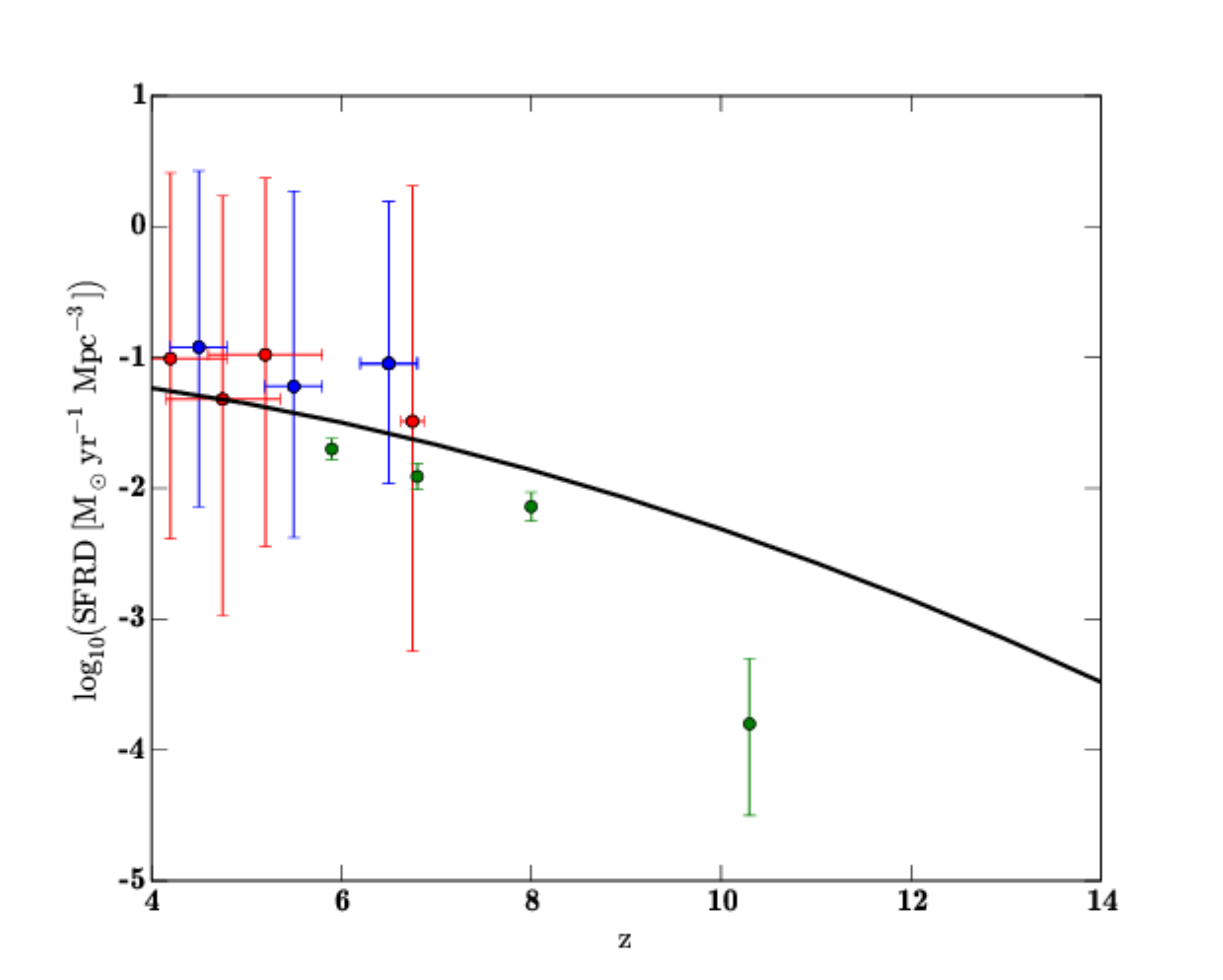} 
\caption{Left panel: Star formation rate versus halo mass for redshifts 6 (red upper lines) and 8 (blue upper lines). The dotted 
lines show the relations taken from the \cite{2011MNRAS.413..101G} galaxies catalogue for low halo masses, the 
dashed lines show the relation taken from the \cite{2007MNRAS.375....2D} galaxies catalogue for high halo masses at 
the same redshifts. The solid lines show the parameterizations from Equation \ref{eq:SFR_param}. Right panel: Star formation rate 
density from the simulations (black solid line), obtained by integrating Equation \ref{eq:SFR_param} 
over the halo mass function. The green circles mark the SFRD corresponding to the UV luminosities corrected 
for dust extinction from \cite{2011arXiv1109.0994B}. The red and blue circles were obtained with measurements of 
gamma-ray bursts by \cite{2013arXiv1305.1630K} and \cite{2012ApJ...744...95R}, respectively.}
\label{fig:SFRD_Mill}
\end{figure*}

The first five observations of star forming galaxies at $z\simeq 6.5$ detected by the ALMA experiment were published 
(see eg. \cite{2013ApJ...773...44W}).  These galaxies have upper limits for the CII luminosity below what is 
predicted by Equation \ref{eq:L_SFR}. However, their SFRs are above ${\rm 10\, M_{\odot}yr^{-1}}$ which puts them in the 
region where a CII deficit was already expected. Observations of typical star forming galaxies at ${\rm z \sim 5-6}$, recently 
obtained with ALMA, show CII luminosities versus FIR ratios clearly above the usual values at z=0 \citep{2015arXiv150307596C}. 
Also, for intensity calculations, according to our model, galaxies with SFRs above ${\rm 10\, M_{\odot}yr^{-1}}$ only represent 
around 20\% of the total CII intensity and so when fitting the CII luminosity versus SFR relation in observational data we should 
take into account that less intense galaxies (which are too faint to be observed especially at high redshifts) have a large weight 
in the CII intensity and that they are more likely to have a more robust $L_{\rm CII}/$SFR ratio.

In order to obtain upper and lower bounds to our CII intensity estimation we decided to use 4 models for the 
$L_{\rm CII}$ versus SFR relation, to which we will refer to as: ${\rm \bf m_1}$, ${\rm \bf m_2}$, ${\rm \bf m_3}$ and ${\rm \bf m_4}$. While 
Equation \ref{eq:L_SFR} corresponds to parameterization ${\rm \bf m_2}$, parameterization ${\rm \bf m_1}$ corresponds to the recent fit to high redshift galaxies by \cite{2014arXiv1402.4075D} and parameterizations ${\rm \bf m_3}$ and ${\rm \bf m_4}$ correspond to fits to the galaxies in Figure \ref{fig:SFR_CII_lum}. These models can all be parameterized as: 
\be
{\rm log10}(L_{\rm CII}[{\rm L_{\odot}}])=a_{\rm LCII} \times {\rm log10}(\psi[{\rm M_{\odot}}])+b_{\rm LCII},
\label{eq:LCII_SFR1}
\ee
with the values for $a_{\rm LCII}$ and $b_{\rm LCII}$ presented in Table \ref{tab:LCII_models}. 
\begin{table}[h]
\centering       
\caption{Parameters for the $L_{\rm CII}$ versus SFR relation}                 
\begin{tabular}{l  c  c  c  c}        
\hline\hline                 
model & $a_{\rm LCII}$ & $b_{\rm LCII}$ \\    
\hline                        
   ${\rm \bf m_1}$  & $0.8475$ & $7.2203$ \\
   ${\rm \bf m_2}$  & $1.0000$ & $6.9647$ \\
   ${\rm \bf m_3}$  & $0.8727$ & $6.7250$   \\
   ${\rm \bf m_4}$  & $0.9231$ & $6.5234$   \\
\hline                                  
\end{tabular}
\label{tab:LCII_models}     
\end{table}

Here the CII intensity was estimated using Equation \ref{eq:Int_CII} with a CII luminosity given by Equation \ref{eq:LCII_SFR1},
converted into a CII luminosity versus halo mass relation. The conversion between SFR and halo mass was made using simulated
galaxy catalogs post-processed by \cite{2007MNRAS.375....2D} and \cite{2011MNRAS.413..101G} from the Millennium and Millennium II dark matter simulations \citep{2005Natur.435..629S,2009MNRAS.398.1150B}. 
We did not use an observational based relation since such a relation is not available for low halo masses and high redshifts. The galaxy SFR from the simulated catalogs is on average related to the mass of the dark matter halo containing the galaxy by:
\be
\psi(M,z)=M_0 \times \left( \frac{M}{M_a}\right)^{a} \left(1 +\frac{M}{M_b}\right)^{b},
\label{eq:SFR_param}
\ee
where the values for the parameters $M_0$, $M_a$, $M_b$, $a$ and $b$ are available in Table \ref{tab:SFR_param} for 
redshifts lower than 20. The evolution of the SFR with mass can be seen in the left panel of Figure \ref{fig:SFRD_Mill} for redshifts 
6 and 8.

\begin{table*}
\centering            
\caption{SFR parameters based in the average relations from the simulated galaxy catalogs}            
\begin{tabular}{l  c c c c c}        
\hline\hline                 
Redshift range & $M_{\rm 0}$ & $M_{\rm a}$ & $M_{\rm b}$ & $a$ & $b$\\    
\hline                 
   0.00-00.50   & $10^{-8.855}$   & $1.0\times10^8$   & $8.0\times10^{11}$  & 2.7  & -4.0 \\
   0.00-02.75   & $10^{-9.097+0.484\times z}$   & $1.0\times10^8$   & $8.0\times10^{11}$  & 2.7  & -4.0 \\
   2.75-03.25   & $3.3\times10^{-8}$   & $1.0\times10^8$   & $4.0\times10^{11}$  & 2.7  & -3.4 \\
   3.50-04.50   & $1.5\times10^{-7}$   & $1.0\times10^8$   & $3.0\times10^{11}$  & 2.6  & -3.1 \\
   4.50-05.50   & $9.0\times10^{-7}$   & $1.0\times10^8$   & $3.0\times10^{11}$  & 2.4  & -2.3 \\
   5.50-06.50   & $3.6\times10^{-6}$   & $1.0\times10^8$   & $2.0\times10^{11}$  & 2.25  & -2.3 \\
   6.50-07.50   & $6.6\times10^{-6}$   & $1.0\times10^8$   & $1.6\times10^{11}$  & 2.25  & -2.3 \\
   7.50-09.00   & $1.0\times10^{-5}$   & $1.0\times10^8$   & $1.7\times10^{11}$  & 2.25  & -2.4 \\
   9.00-11.00   & $3.7\times10^{-5}$   & $1.0\times10^8$   & $1.7\times10^{11}$  & 2.1  & -2.2 \\
  11.00-13.00   & $5.0\times10^{-5}$   & $1.0\times10^8$   & $1.5\times10^{11}$  & 2.1  & -2.2 \\
  13.00-20.00   & $[5.0+(z-13.0)]\times10^{-5}$  & $1.0\times10^8$ & $[1.5+(z-13)\times0.015]\times10^{11}$  & 2.1  & $-2.2-(z-13)\times0.03$\\
  \hline                                  
\end{tabular}
\label{tab:SFR_param}     
\end{table*}

The use of this formula results in the star formation rate density (SFRD) evolution shown in the right panel of Figure \ref{fig:SFRD_Mill} assuming a dark matter halo mass range from $10^8\, {\rm M_{\odot}}$ to $10^{14}\, {\rm M_{\odot}}$. 
The Millennium and Millennium II simulations only goes till a redshift of $20$. However unless we want to consider unusual stars 
the relation for $z\, =\,20$ should be a good approximation for $z\, >\, 20$, if required.

\subsection{Calculating CII emission using gas physics}

\begin{figure}[htbp]
\hspace{-4mm}
\includegraphics[scale = 0.48]{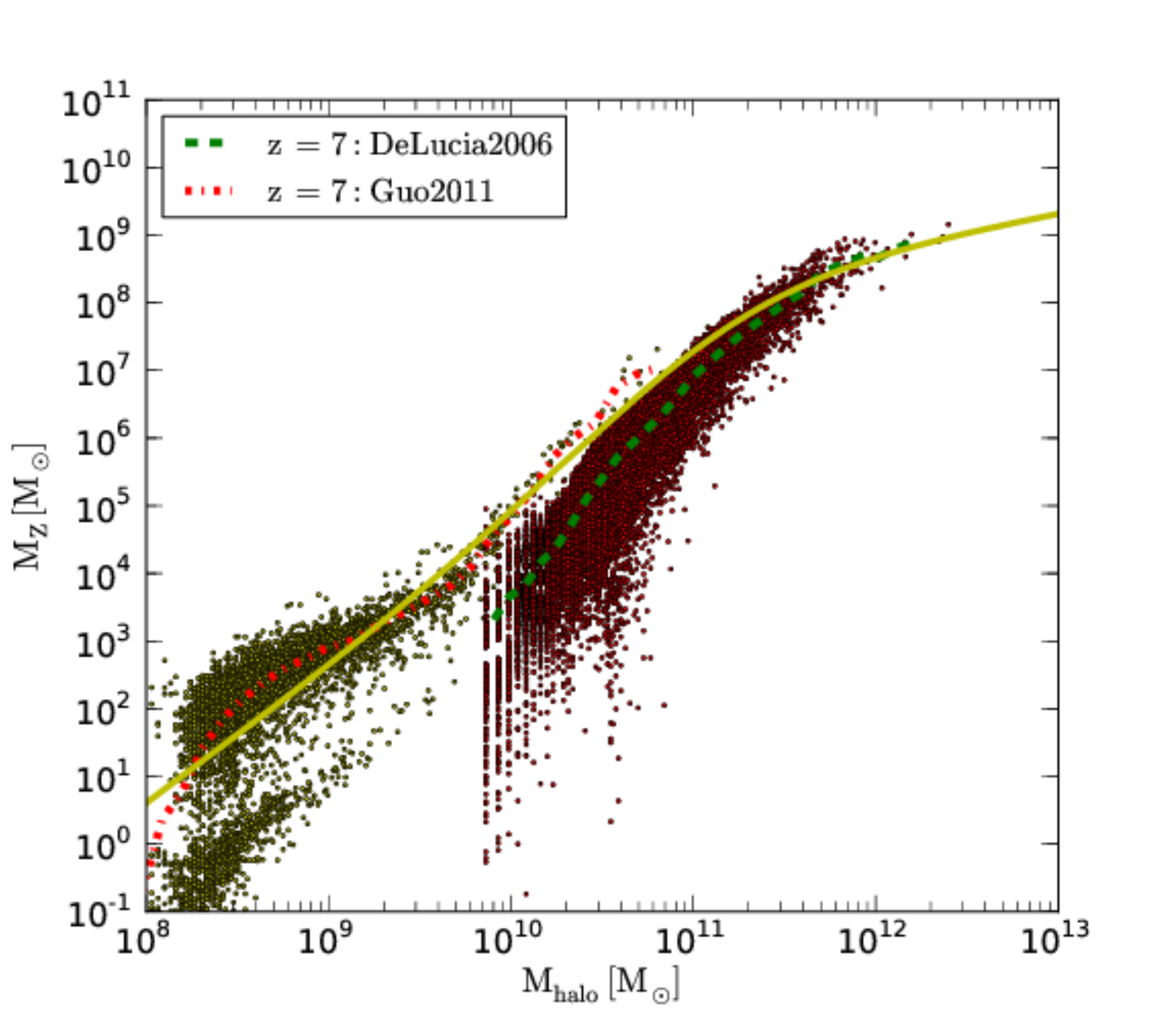} 
\caption{Mass in metals in the hot gas component $M_{\rm Z}$ as a function of the halo mass $M$ at $z\approx7$. The 
dahed dotted line shows the mean 
relation from the \citep{2011MNRAS.413..101G} galaxy catalog, the dashed line shows the mean relation from the 
\citep{2007MNRAS.375....2D} galaxy catalog and the solid line shows our fits to the mean values. The dots show 
the dark matter halos in the two catalogs.}
\label{fig:Metallicity}
\end{figure}

The maximum possible upper value for the CII emission can be obtained assuming that all the carbon in the hot gas 
(typical HII regions) in a galaxy is ionized and therefore emitting in the CII line, such as was done in 
\cite{2012ApJ...745...49G}.
Here, we do a similar calculation but with an improved parameterization of the metallicity in the galaxies hot gas 
obtained using the \cite{2011MNRAS.413..101G} galaxies catalog for low mass halos and the \cite{2007MNRAS.375....2D} 
galaxies catalog for high halo masses. 
The resulting relation between halo mass and metallicity in the hot gas component is shown in Figure \ref{fig:Metallicity}. By 
analysing this figure we found that the metallicity in the lower mass halos of the \cite{2007MNRAS.375....2D} 
simulation is lower than the one found in the halos from the \cite{2011MNRAS.413..101G} simulation, although these 
simulations used very similar prescriptions to determinate the galaxies metallicity.  Since the \cite{2011MNRAS.413..101G} 
simulation has a much higher mass resolution, we believe that their results are more reliable for the low luminosity halos, since the halos in the \cite{2007MNRAS.375....2D} galaxies catalog are only well resolved for masses above $3\times 10^{\rm 10}\, {\rm M_{\odot}}$. 

The average relation between $M_Z$ and halo mass $M$ in the referred simulated galaxy catalogs can therefore, be parameterized in the form:
\ba
M_Z(M) &=& M_0 \left(M/M_{\rm a}\right)^{\rm a} \left(1+M/M_{\rm b}\right)^{\rm b} \\ \nonumber
       &\times& \left(1 + M/M_{\rm c} \right)^{\rm c}\times \left(1 + M/M_{\rm d} \right)^{\rm d}, 
\ea
where at the redshift range 5.0 to 8.5 these parameters take the values: 
$M_{\rm 0}\, =\, z-1$, $M_a\, =\, 1.0\times\,10^8\ {\rm M_{\odot}}$, $M_b\, =\, 9.0\, \times\, 10^{9}\ {\rm M_{\odot}}$, 
$M_c\, =\, 2.0\, \times\, 10^{12}\, {\rm M_{\odot}}$, $M_d\, =\, 2.0\, \times\, 10^{13}\,  {\rm M_{\odot}}$, $a=1.7$, $b=1.0$, $c=-5.0$ and $d=2.5$. 

Assuming that all the carbon in the hot gas is ionized and that the carbon mass corresponds to a fraction of 
21\% of the total mass in metals (this is the same percentage of carbon found in the sun) we obtain 
$M_{\rm CII}\, =\, 0.21\, M_{\rm Z}$. In reality only a fraction of the carbon in HII regions is ionized which can be easily included 
in these calculations.
At large enough volumes the number density of CII atoms can be estimated from the halos mass using the formula:
\be
n_{\rm CII}^{\rm sim}(z)=\int^{M_{\rm max}} _{M_{\rm min}} dM \frac{dn}{dM} \frac{M_{\rm CII}(M,z)}{m_{\rm c}},
\label{eq:2}
\ee
where $m_{\rm c}$ is the atomic carbon mass.

We can obtain an upper value for the intensity of CII emission in HII regions by replacing in Equation \ref{eq:ICII_teorico} the 
CII number density obtained from Equation \ref{eq:2}. We estimated the CII number density by assuming that HII regions have an 
electronic temperature of $10^4$ K and an electronic density of $10^4\, \rm cm^{-3}$ (these values correspond to saturation emission values as obtained in \cite{2012ApJ...745...49G}).   
\begin{figure}[htp]
\hspace{-6mm}
\includegraphics[scale = 0.49]{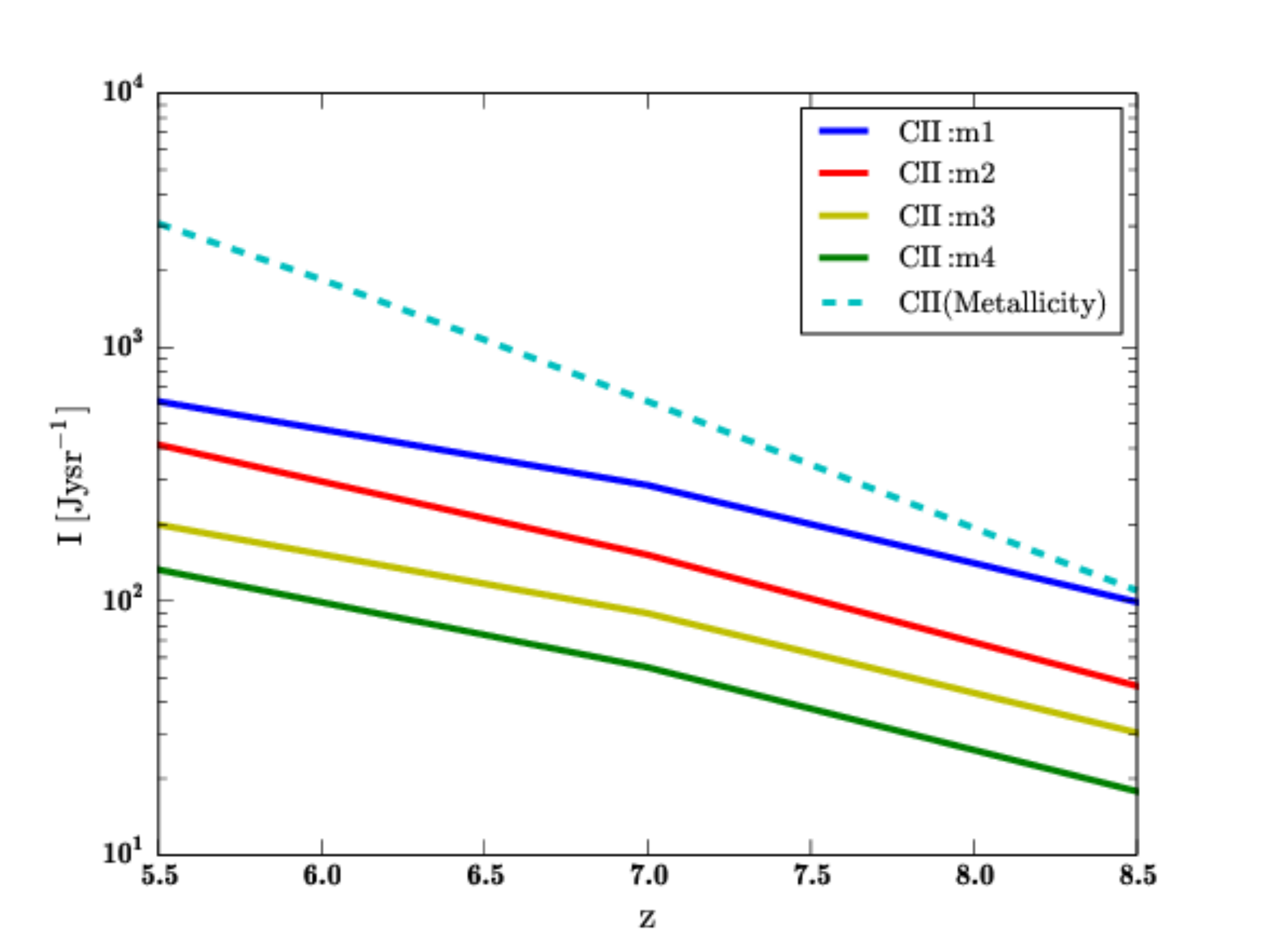} 
\caption{CII intensity as a function of redshift according to  the CII models ${\bf m1}$, ${\bf m2}$, ${\bf m3}$ and ${\bf m4}$ (upper to lower 
solid lines). The cyan dashed 
line corresponds to the CII luminosity from HII regions (assuming that all the carbon is ionized).}
\label{fig:Int_CII_varios}
\end{figure}

In Figure \ref{fig:Int_CII_varios} we show the CII intensity estimated assuming several models for the CII emission. The average 
intensity of CII emission in the redshift range shown is between 
$\bar{I}_{\rm CII}\, \approx\, 6\times10^{2}\, {\rm Jy\ sr^{-1}}$ for model ${\rm \bf m_1}$ and  
$\bar{I}_{\rm CII}\, \approx\, 9\times10^{1}\, {\rm Jy\ sr^{-1}}$ for model ${\rm \bf m_4}$. The average CII intensity, obtained by averaging 
models ${\rm \bf m_1}$ to ${\rm \bf m_4}$, between $z\sim5.5$ and $z\sim8.5$ is $\bar{I}_{\rm CII}\, \approx\, 4\, \pm 2 \times10^{2}\, {\rm Jy\ sr^{-1}}$.

\begin{figure}[htbp]
\hspace{-5mm}
\includegraphics[scale = 0.49]{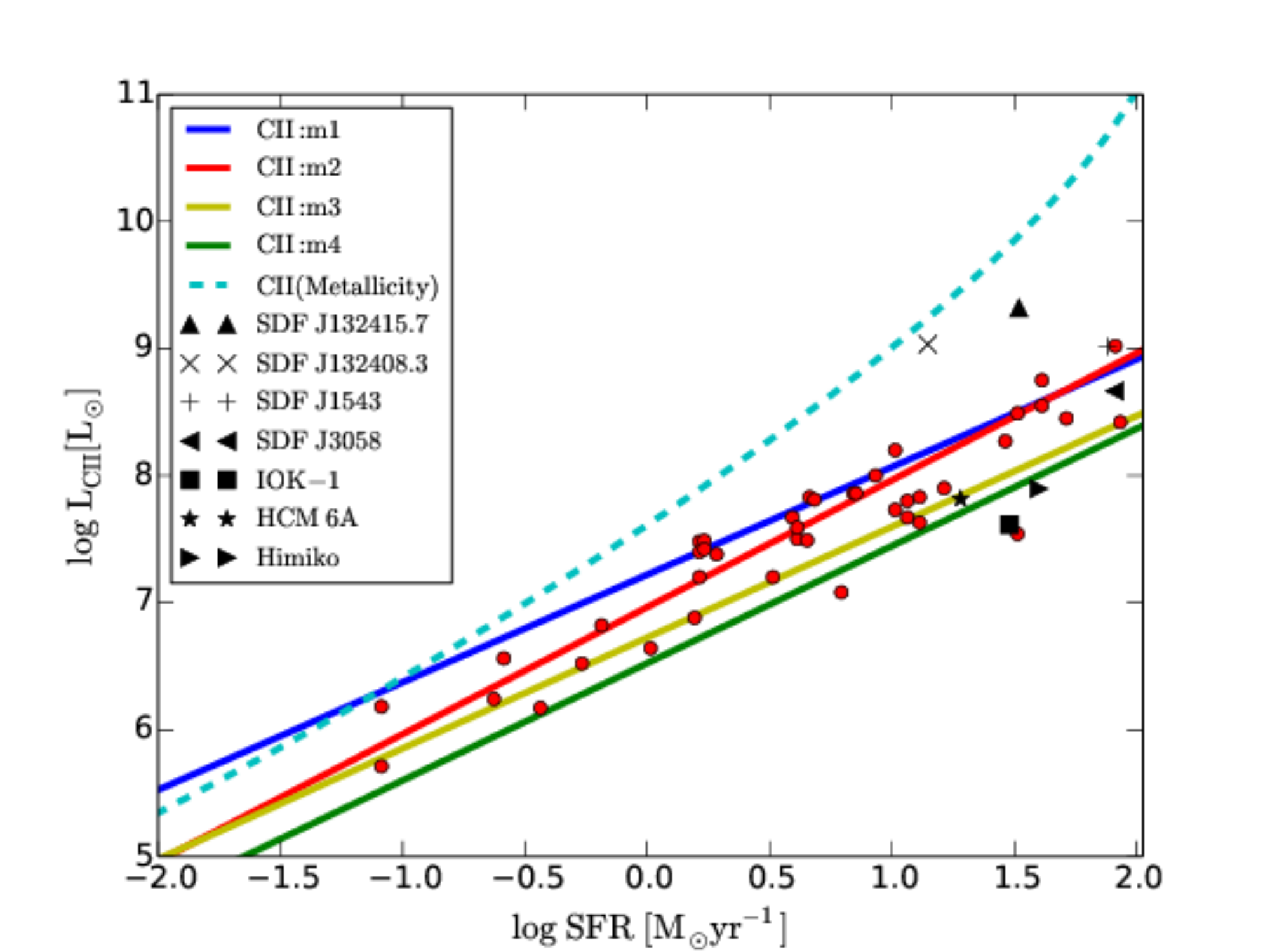} 
\caption{CII luminosity as a function of the SFR. The cyan dashed 
line corresponds to the CII luminosity from HII regions (assuming that all the carbon is ionized) and the solid 
lines corresponds to the CII luminosity obtained from the SFR using relations ${\bf m1}$, ${\bf m2}$, ${\bf m3}$ and ${\bf m4}$ 
(upper to lower lines). 
The red dots are local universe galaxies from the ISO Key Program \citep{2001ApJ...561..766M} and 
the other symbols are upper limits from galaxies at $z\, >\, 6.5$
(taken from: \citep{2014arXiv1401.3228G,2013ApJ...771L..20K,2014arXiv1405.5387O}).}
\label{fig:SFR_CII_lum}
\end{figure}

In Figure \ref{fig:SFR_CII_lum}, the CII luminosity as a function of the SFR for the different methods 
described, is shown together with observational points of normal local galaxies from \cite{2001ApJ...561..766M} 
and with observational upper limits for high redshift galaxies. The 
observed high redshift galaxies, presented in this figure, have high SFRs which indicates that they are
very massive and rare or that they have extreme SFRs/Mass ratios. In either 
case these galaxies have little effect on the overall CII intensity. 

The CII luminosities from ionized regions, presented in Figure \ref{fig:SFR_CII_lum}, 
were obtained by assuming that $L_{\rm CII}$ is linearly proportional to the halo mass and by determining the constant of 
proportionality between the two by imposing that Equations \ref{eq:Int_CII} and \ref{eq:ICII_teorico} give the same result. The 
relation between halo mass and SFR was assumed to follow Equation \ref{fig:SFR_CII_lum}.

\section{CII Foregrounds}

The CII line emitted in the redshift range z $\approx$ 8.51 - 5.35 is observed at frequencies 
${\rm 200\, -\, 300\, GHz}$. CII is a far-infrared line and so CII intensity maps will be contaminated by other infrared lines 
and by infrared continuum emission from galaxies and from the IGM. In this section we show estimations for the contamination from all of these extra-galactic sources in the relevant observing frequency band. In addition we also consider contamination due to emission from our galaxy.

\subsection{Contamination from line emission}

The main contaminants in CII intensity maps will be emission lines from lower redshifts namely the ${\rm [OI]\, 145\, \mu m}$, the 
${\rm [NII]\, 122\, \mu m}$, 
the ${\rm [NII]\, 205\, \mu m}$ and the CO rotation lines from transitions CO(2-1) and higher.\\
The ${\rm [OI]\, 145\, \mu m}$ and the ${\rm [NII]\, 205\, \mu m}$ lines are typical of PDRs while the ${\rm[NII]\, 122\, \mu m}$ line 
is typical of HII regions and so the SFR can be used to roughly estimate their intensity of emission such as in the CII case (see: section \ref{sec:other_contamination_lines}). 
The CO lines are emitted from molecular gas and their luminosities depend on several characteristics of the gas 
and so we carefully estimate their intensity of emission in the next section.

\subsection{CO signal from simulations}
\label{sec:CO}

CO rotation lines will be the main contaminants in CII intensity maps observed at 
frequencies 200 $-$ 300 GHz. Since the luminosities of the several relevant CO transitions are poorly constrained observationally, we estimate
their intensities using the CO fluxes in the simulated galaxy catalog from \citet{2009ApJ...702.1321O} and confirm 
our results with a CO intensity calculated using only observational relations, when available.
The \citet{2009ApJ...702.1321O} catalog is available for halo masses above $10^{10}\,{\rm M_{\rm \odot}}$ and  provides astrophysical properties such as the CO fluxes for rotational transitions (1-0) to (10-9). The CO emission was estimated from the galaxies molecular gas content and from the ISM temperature using physically based prescriptions 
and assuming thermal equilibrium.

Each CO rotation line that is observed in the frequency range 200 - 300 GHz will come from the redshift range shown 
in Table \ref{tab:CO}. Note that for the CO(2-1), as is shown in Table \ref{tab:CO}, the minimum relevant redshift for this study 
is zero which corresponds to the line rest frequency. 
\begin{table}[h]
\centering    
\caption{Emission redshifts for the CO rotational transitions in the frequency range 200 - 300 GHz}                   
\begin{tabular}{l  c c c}        
\hline\hline                 
transition (J) & $\nu_{\rm CO}^{\rm J}\, ({\rm GHz})$ & ${\rm z}(\nu_o\, \approx\, {\rm 300\, GHz})$ & ${\rm z}(\nu_{\rm o} \approx {\rm 200\, GHz})$ \\    
\hline                        
   2-1   & 230.542   & $0$  & $0.150$ \\
   3-2   & 345.813   & $0.150$ & $0.730$  \\
   4-3   & 461.084   & $0.535$ & $1.305$ \\
   5-4   & 576.355   & $0.920$ & $1.881$ \\
   6-5   & 691.626   & $1.305$ & $2.458$ \\
   7-6   & 806.897   & $1.690$ & $3.035$\\
   8-7   & 922.168   & $2.074$ & $3.610$ \\
   9-8   & 1037.439  & $2.458$ & $4.186$ \\
   10-9  & 1152.71   & $2.842$ & $4.762$ \\
\hline                                  
\end{tabular}
\label{tab:CO}     
\end{table}

\begin{table*}
\centering     
\caption{CO transitions luminosity parameters}                  
\begin{tabular}{l  c c c c c c c c c c}       
\hline\hline                 
 Trans & $L_0$ & $M_{c1}$ & $M_{c2}$ & $M_{c3}$ & $M_{c4}$ & $d0$ & $d1$ & $d2$ & $d3$ & $d4$  \\    
\hline                        
   2-1   & $4.70\times10^{-29}$  & $1.0\times10^{11}$ & $6.0\times10^{11}$ & $5.0\times10^{12}$ & $5.0\times10^{14}$   & $3.05$& $-2.0$  & $-2.3$  & $1.9$  & $5.0$\\
   3-2   & $3.00\times10^{-24}$  & $6.0\times10^{11}$ & $5.0\times10^{12}$  & $4.0\times10^{13}$ & $0.0$   & $2.6$& $-3.5$  & $0.2$ & $2.2$  & $0.0$ \\
   4-3   & $8.00\times10^{-18}$  & $9.0\times10^{11}$ & $5.0\times10^{12}$  & $3.0\times10^{13}$ & $0.0$   & $2.05$& $-1.7$  & $-1.8$ & $2.3$  & $0.0$ \\
   5-4   & $3.50\times10^{-18}$  & $2.0\times10^{12}$ & $4.0\times10^{12}$ & $1.0\times10^{13}$ & $0.0$   & $2.05$& $-2.0$  & $-3.9$ & $4.5$  & $0.0$ \\
   6-5   & $4.00\times10^{-18}$  & $9.0\times10^{10}$ & $6.0\times10^{11}$ & $0.0$ & $0.0$   & $2.0$& $1.5$  & $-3.75$ & $0.0$  & $0.0$ \\
\hline                                  
\end{tabular}
\label{tab:CO_param_M}     
\end{table*}

The CO intensity can be estimated from its luminosity as:
\be
\bar{I}_{\rm CO}(\nu) =  \sum_J \int_{M_{\rm min}}^{\infty} dM \frac{dn}{dM}(z,M)\frac{L^{\rm J}_{\rm CO}(z,M)}{4\pi D_{\rm L}^2}y^J(z)D_{\rm A}^2,
\label{eq:I_ave}
\ee
\\
where the sum in J (angular momentum) is a sum over the luminosities of the different rotation lines from CO(2-1) to CO(6-5)  
and $z\, =\, \nu_{\rm CO}^J/\nu_{\rm o}-1$.
We do not account for higher CO transitions since according to this CO model the CO contamination in CII intensity maps is highly dominated
by the lower CO transitions.
We also justify
our choice by arguing that the contamination from transitions (7 - 6) and higher is originated from high 
redshifts $(4.8\, >\, z\, >\, 1.7)$ and so the lower metallicity of these galaxies is likely to result in a considerably low CO emission.

Using the simulated fluxes we parameterized the CO luminosity of galaxies as a function of halo mass for transitions 
CO(2-1) to CO(6-5) as:
\ba
\label{eq:CO_Luminosity0}
L^{\rm J}_{\rm CO}({M})[L_{\odot}]&=&L_0\times M^{\rm d0}\times \left( 1+\frac{ M}{ M_{\rm c1}}\right)^{\rm d1}\\ \nonumber
        &\times&\left( 1+\frac{ M}{ M_{\rm c2}}\right)^{\rm d2} \left( 1+\frac{ M}{ M_{\rm c3}}\right)^{\rm d3}\left( 1+\frac{ M}{ M_{\rm c4}}\right)^{\rm d4},
\ea
\\
where the parameters for each transition can be found in Table \ref{tab:CO_param_M}. Note that these parameters were obtained for each transition 
by averaging the CO luminosity in the redshift range shown in Table \ref{tab:CO}.

\begin{figure}[htp]
\hspace{-4mm}
\includegraphics[scale = 0.40]{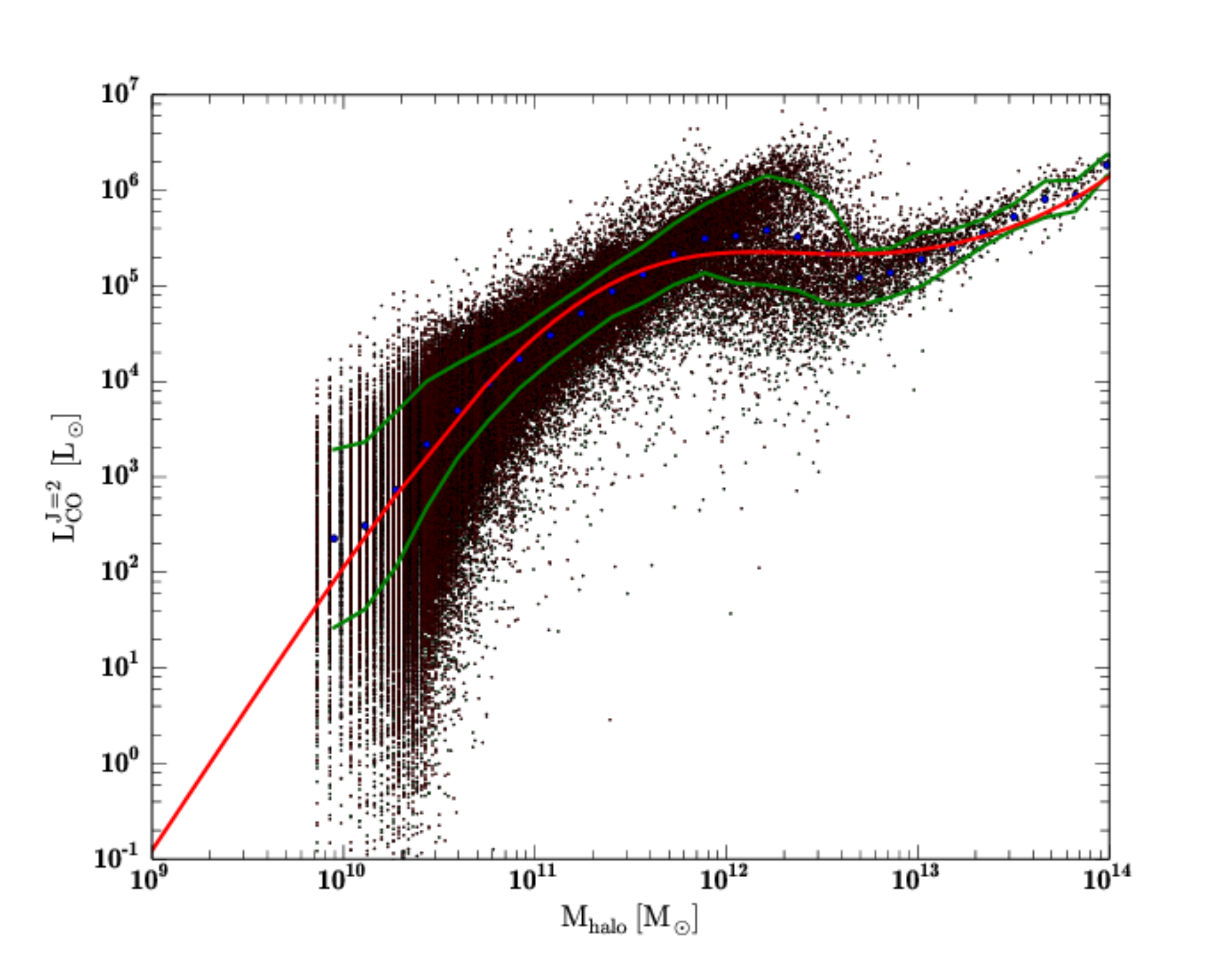} 
\caption{Luminosity of the CO(2-1) transition as a function of halo mass for the redshift band $z\, \sim\, 0-1.5$. The black dots correspond to halos 
in the \citet{2009ApJ...702.1321O} simulation. The blue dots show the mean of the scatter when binned in 30 logarithmic intervals in mass. The solid green 
lines show the $\pm 1\sigma$ relation. The red solid line corresponds to the parameterization given by Equation \ref{eq:CO_Luminosity0} with the 
parameters from Table \ref{tab:CO_param_M}.}
\label{fig:LCO_J2}
\end{figure}

Figure \ref{fig:LCO_J2} shows the luminosity of the CO(2-1) transition as a function of halo mass, in the redshift range 0 to 0.15, obtained from the simulation.

Given that the minimum halo mass available in the \cite{2009ApJ...702.1321O} galaxy catalog is not low enough for our study, we extrapolated the average CO luminosity 
to lower halo masses assuming that it is proportional to SFR at low masses.
The CO luminosities were parameterized as a function of dark matter halo masses and not as a function of galaxy masses 
and therefore for high halo masses they include the contribution from a main galaxy and several satellite galaxies.
This parameterization could also have been made as a function of IR luminosity or SFR, however, galaxies powered by active
galactic nuclei have relatively small IR luminosities, small SFRs and high CO fluxes and so this population 
would have to be taken into account separately.

\begin{figure*}[!t]
\hspace{-5mm}
\includegraphics[scale = 0.54]{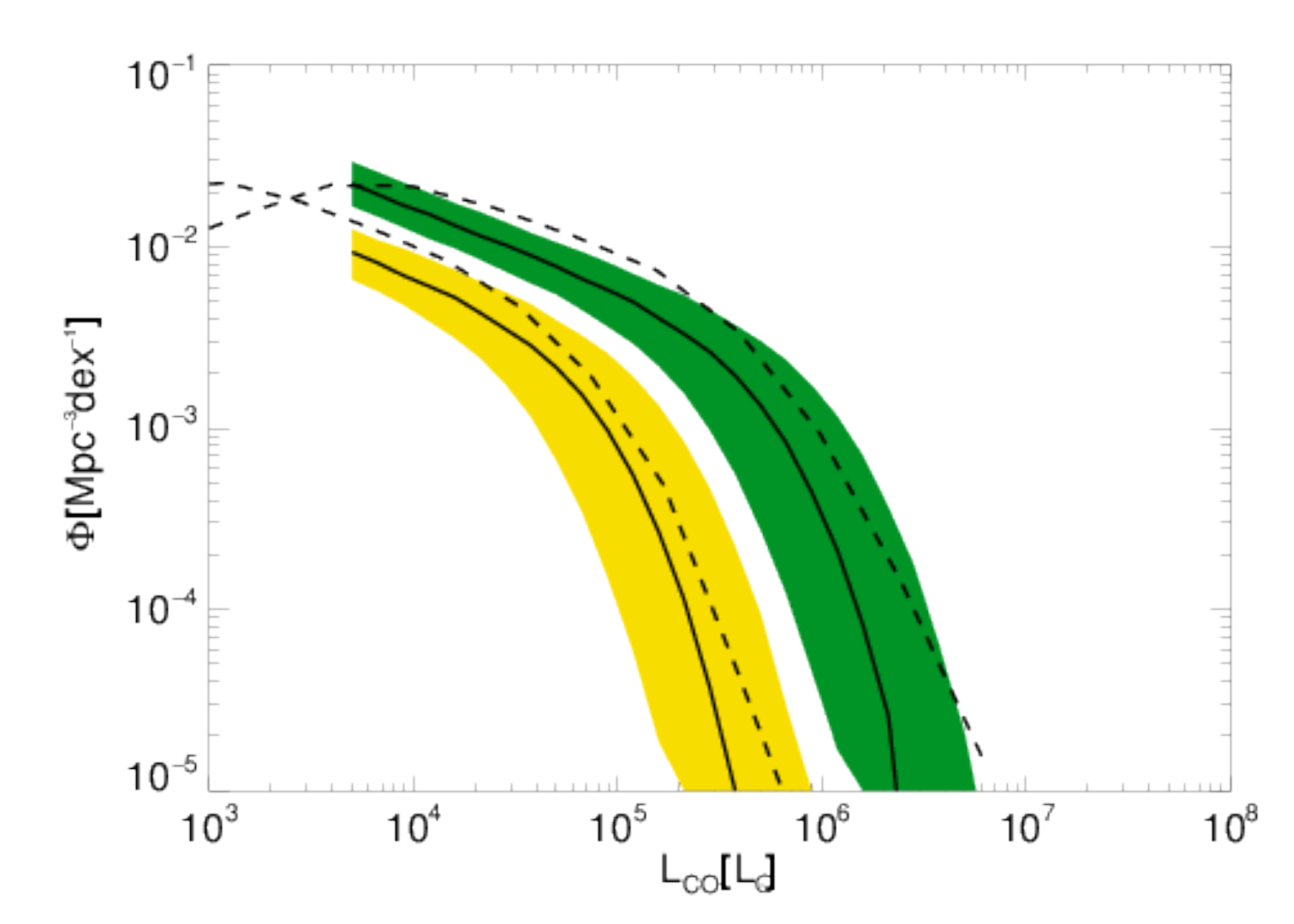}
\hspace{-5mm}
\includegraphics[scale = 0.54]{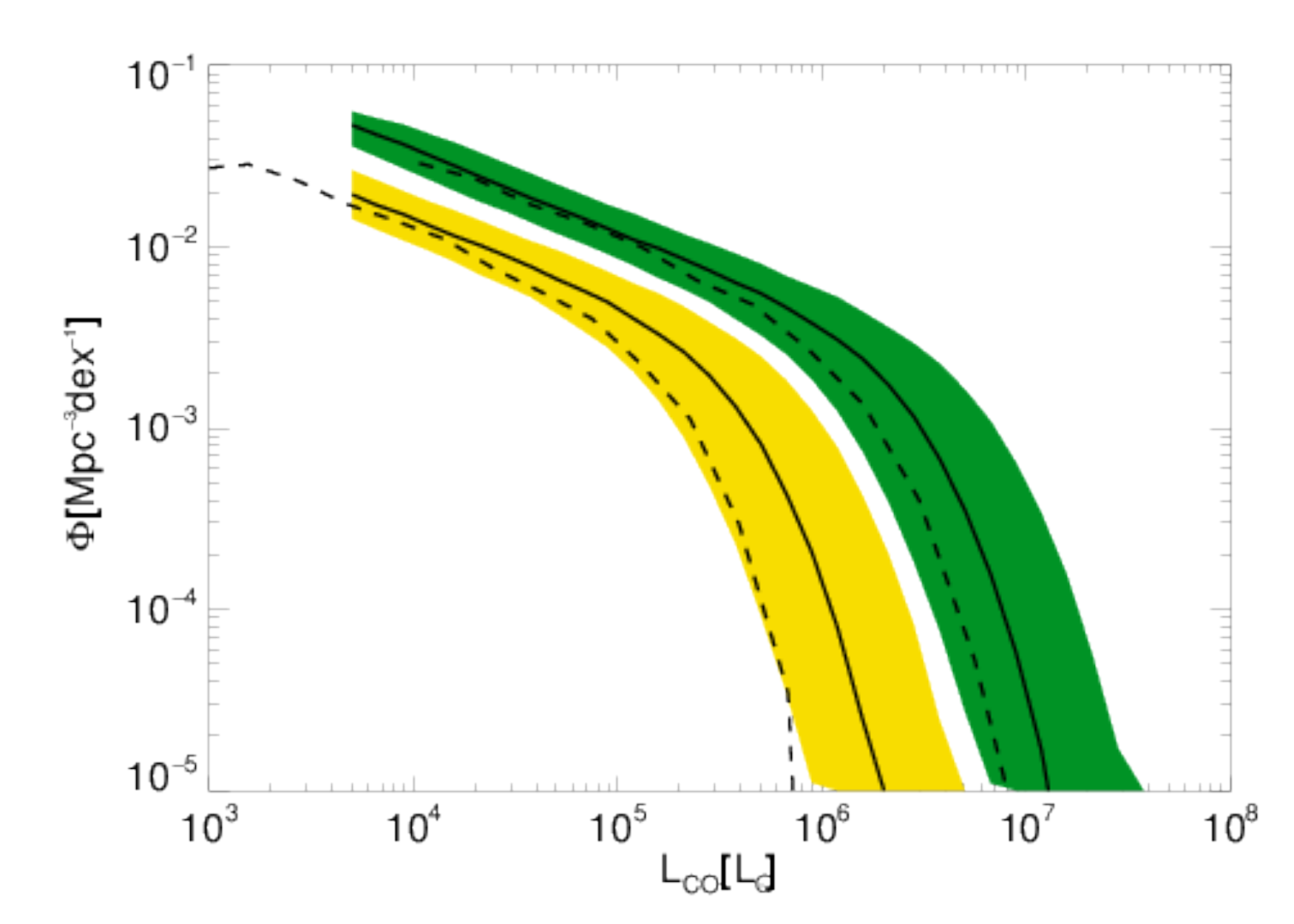}
\caption{ \label{fig:CO_LF} CO luminosity functions based on the \cite{2009ApJ...702.1321O} CO model (dashed lines) and 
on the observational CO model (solid lines) at redshifts 
0 (left panel) and 1 (right panel). The yellow and green regions show the uncertainty in the CO observational model due only
to the uncertainty in the conversion factor between IR luminosity to CO(1-0) luminosity. The upper curves correspond to the 
transition CO(2-1) and the lower curves correspond to transition CO(1-0)}
\end{figure*}

The theoretical average power spectra of CO contamination presented in Figure \ref{fig:ps_3D_CO_Jb} for transitions 
CO(2-1) to CO(6-5) indicates that the dominant contamination will be due to the 
low-J CO transitions. However, the ratios of different CO lines were obtained by assuming a simple model with a single gas 
component in local thermodynamic equilibrium. This does not necessarily have to represent well the molecular gas conditions in all types of galaxies. A more recent work described in \cite{2011MNRAS.418.1649L} and \cite{2012MNRAS.426.2142L},  attempts to estimate the luminosity of the several CO transitions using an improved method to estimate the molecular gas content in galaxies. This is based in a somewhat more detailed model of the gas properties, as compared to \citet{2009ApJ...702.1321O}, used to estimate the relation between CO emission and molecular gas content.
The main difference in the results obtained by these two authors is that the \cite{2012MNRAS.426.2142L} model 
predicts a smaller molecular content
in galaxies for $z\, >\, 2$ and higher ratios between the CO luminosities for higher transitions. 
These two corrections practically compensate themselves in terms of contamination in CII maps at the relevant 
frequencies for these study and so they should not have a significant effect in the validity of our predictions for 
intensity mapping. Even though there are limitations to the CO luminosities calculation made by \cite{2009ApJ...702.1321O}, observationally 
only the CO(1-0) line is well constrain at small redshifts ($z\, \le\, 1$) and in that case the CO LFs derived from the simulated 
galaxies catalog are compatible with observations. The few CO observations at $z\, >\, 1$  suggest a number density of CO 
emitters higher than what is predicted by the Obreschkow model \citep{2010ApJ...713..686D,2010Natur.463..781T,2012MNRAS.426..258A}, 
however these observations are restricted to mainly the CO(2-1) transition from very high luminosity galaxies while for the relevant 
observed frequency range the CO emission is originated at $z\, <\, 1$, where the models are in better agreement.

\subsection{CO signal from observations}
\label{sec:CO_o}

An observational only based model to estimate the CO(1-0) luminosity is presented in \cite{2013arXiv1303.4392S} and so to support our conclusions we used this completely independent model to estimate the CO contamination in CII maps. The Sargent model estimates CO emission from a recent IR luminosity function at $z\, =\, 1$ presented in \cite{2012ApJ...747L..31S} and with the observational relations between IR and CO luminosities presented in \cite{2013arXiv1303.4392S}.
The luminosity function (LF) is an useful way to put constraints on the overall luminosity of observed galaxies above a given 
luminosity limit characteristic of each survey. It corresponds to the number density of galaxies per luminosity interval as a function of luminosity. The LF is commonly plotted in units of number density per decade in luminosity $({\rm \Phi[Mpc^{-3}\, dex^{-1}]}$,  where dex accounts for the logarithmic variation of the luminosity for the bin used
(${\rm log_{\rm 10}}\, L_{\rm f}\, -\, {\rm log}_{\rm 10}\, L_{\rm i}$).
{The Sargent IR LF for $z\not=1$ can be scaled with redshift using the factor $(1+z)^{2.8}$ in the galaxies luminosity and 
scaling the number density as $\Phi\, \propto\, (1+z)^{-2.4}$ for $z\, >\, 1.0$ (for lower redshifts the number density is fixed).

The CO luminosities can be obtained from the IR LF using:
\be
{\rm log} \left(\frac{L^{ \prime }_{\rm CO(J=1\to0)}}{\rm K\ km\ s^{-1}\ pc^2 }\right)= \alpha_1+\beta_1 {\rm log} \left( \frac{L_{\rm IR}}{\rm L_{\odot}} \right)
\ee
\\
where $(\alpha_1,\, \beta_1)=(0.18\, \pm\, 0.02;\, 0.84\, \pm\, 0.03)$ for normal galaxies and 
$(\alpha_1,\, \beta_1)\,=\, (-0.28^{+0.15}_{-0.09};\, 0.84\, \pm\, 0.03)$ for starbursts \citep{2013arXiv1303.4392S}.
In Figure \ref{fig:CO_LF} the CO LF based in the \cite{2009ApJ...703.1890O} model was obtained using a halo mass function and the CO luminosity parameterization from Equation \ref{eq:CO_Luminosity0}. The two models shown in Figure \ref{fig:CO_LF} agree taking 
into account the uncertainty in the relation between the IR luminosity and CO(1-0) luminosity used in the observational CO model (showed as shaded regions). The uncertainties in the Sargent CO LFs are even higher if we take into account the error bars in 
the IR luminosity function, or the uncertainty in the passage from the CO(1-0) line to higher transitions. 

Ratios between the luminosities of the CO(1-0) line and higher CO transitions (in units proportional to the surface 
brightness $L'_{\rm CO}\, [{\rm K\, km\, s^{-1}\, pc^2}]$) for 
different types of galaxies are available in \cite{2013ARA&A..51..105C}. In order 
to obtain observationally only based estimations for the LFs of the relevant CO transitions we used the Sargent 
CO(1-0) model plus the \cite{2013ARA&A..51..105C}  observational ratios: ${\rm r_{21}\, =\, 0.85}$, ${\rm r_{31}\, =\, 0.66}$, ${\rm r_{41}\, =\,0.46}$ 
and ${\rm r_{51}\, =\, 0.39}$ which are appropriate for submillimeter galaxies. Since there is no 
available observational relation between transitions CO(6-5) and CO(1-0), we assumed that ${\rm r_{61}\, =\, r_{51}}$. 
For each transition the CO luminosity in [${\rm L_{\odot}}$] can be obtained using:
\be
L_{\rm CO}\, =\, 3\times10^{-11}\, \nu^3\, L'_{\rm CO}.
\ee
\begin{figure}[htbp]
\hspace{-11pt}
\includegraphics[scale = 0.48]{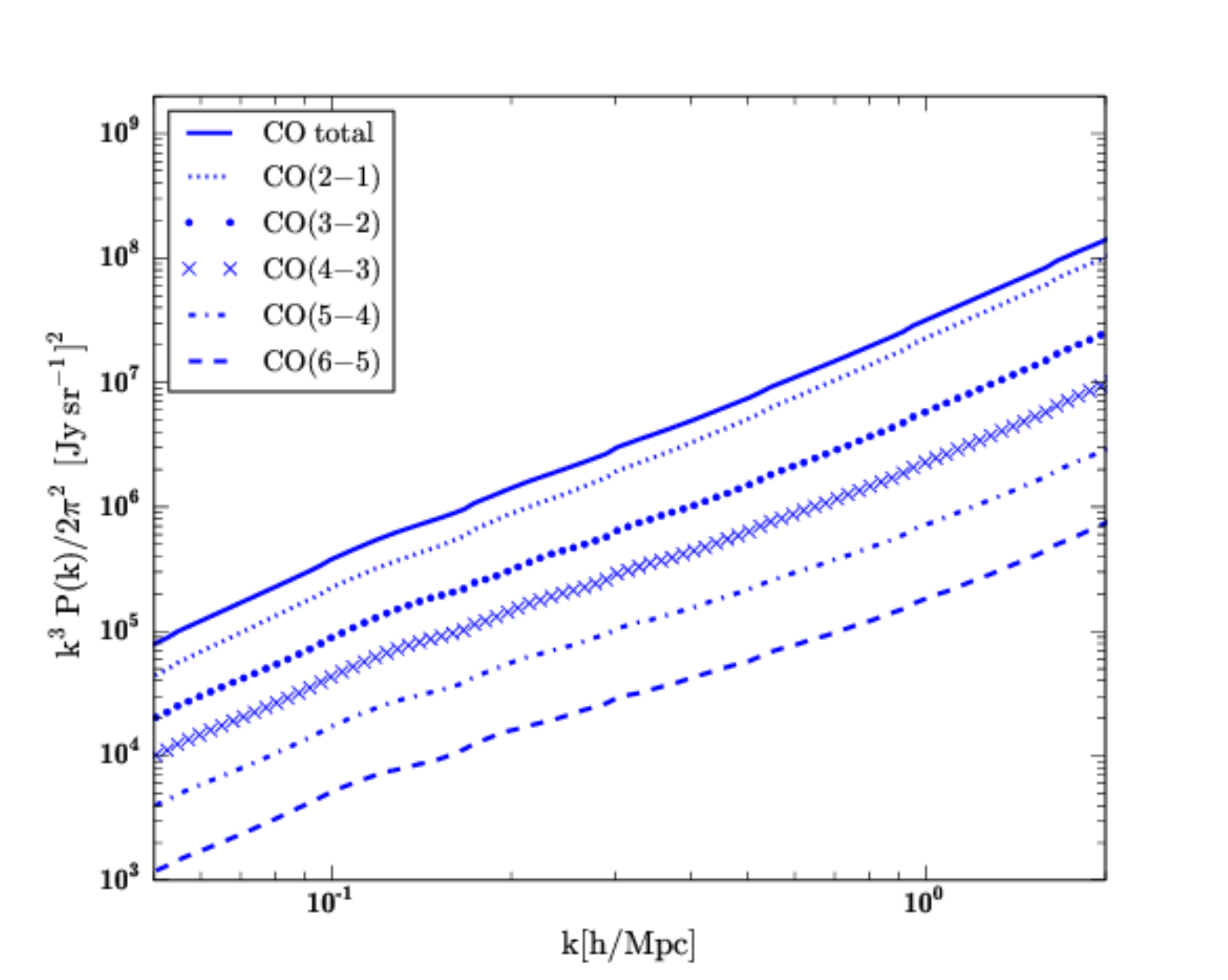} 
\caption{Power spectra of CO line emission made using the parameterization from Equation \ref{eq:CO_Luminosity0} in the frequency range 200 GHz to 300 GHz projected to the redshift of the CII line. The top solid line shows the 
total CO power spectra and the lower lines show the CO(2-1) to CO(6-5) transitions.}
\label{fig:ps_3D_CO_Jb}
\end{figure}

\begin{figure}[htbp]
\hspace{-11pt}
\includegraphics[scale = 0.48]{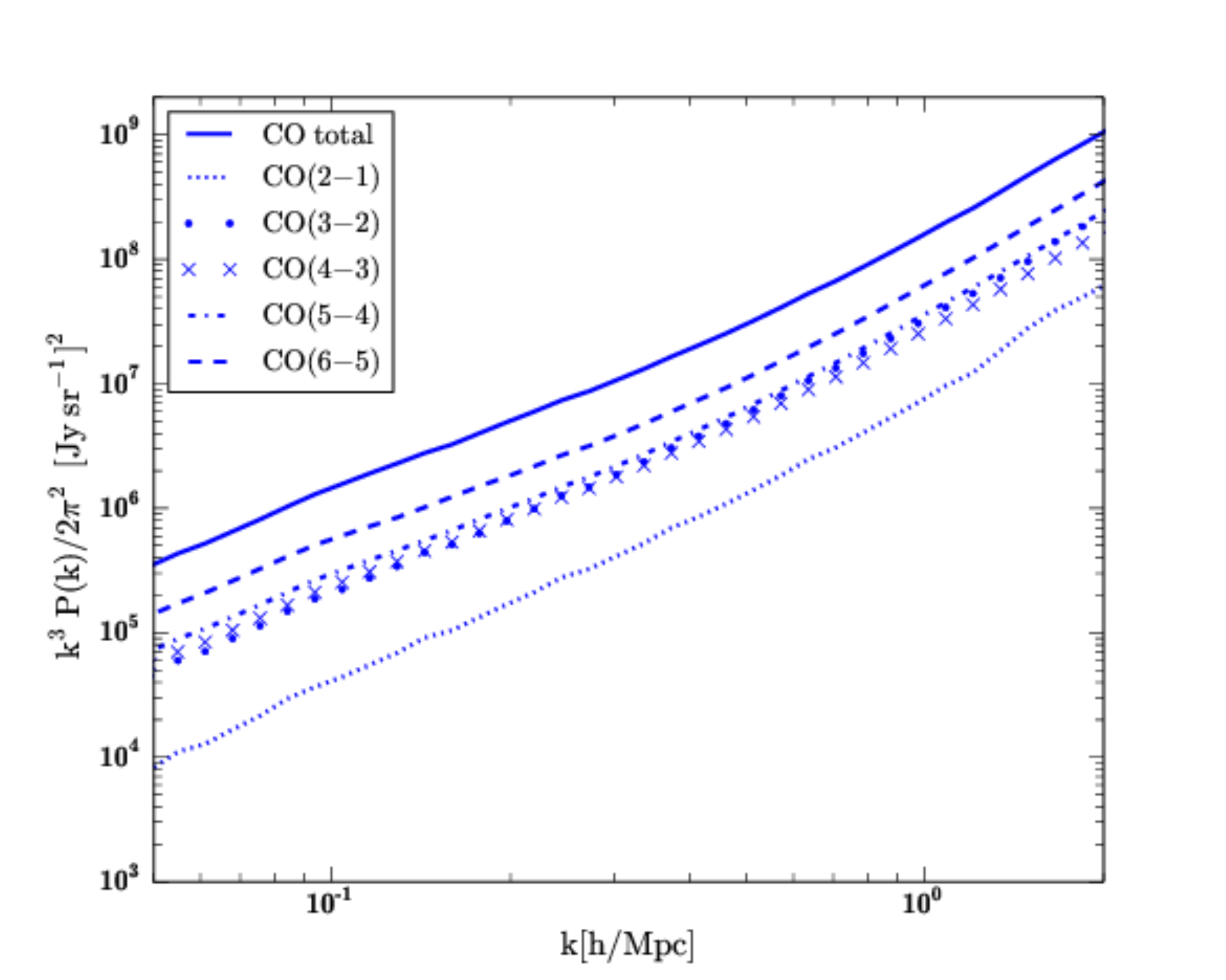} 
\caption{Power spectra of CO line emission made using the observational CO model in the frequency range 200 GHz to 
300 GHz projected to the redshift of the CII line. The top solid line shows the total CO power spectra and the lower 
lines show the CO(2-1) to CO(6-5) transitions.}
\label{fig:ps_3D_CO_Jc}
\end{figure}

We will from now on refer to the observational CO luminosities predicted using the Sargent CO(1-0) LF plus the Carilli 
et al. ratios for CO transitions as the observational CO model.
The main differences between the two CO models lies in the conversion 
ratios between the luminosity of the several transitions given that the average ratio in the Obreschkow simulation are 
${\rm r_{\rm 21}\, =\, 0.93}$, ${\rm r_{\rm 31}\, =\, 0.70}$, ${\rm r_{\rm 41}\, =\, 0.38}$, ${\rm r_{\rm 51}\, =\, 0.12}$ 
and ${\rm r_{\rm 61}\, =\, 0.02}$ 
at a redshift close to zero and slightly increase for higher redshifts. The flux ratios in the Obreschkow simulation
are appropriate for regular galaxies, while for star bursts and quasars the ratios between fluxes of high CO transitions are much higher. 
Recent observationally based ratios for different CO transitions as a function of redshift can be found in \cite{2014arXiv1409.8158D}. This study suggests 
that the relative contribution from high CO transitions relevant for our study is even smaller than what is predicted by the two models discussed here.  That is, it should be easier to remove CO contamination from observational maps.
Given the lack of observational measurements of fluxes of high CO transitions in normal galaxies, with masses below ${\rm 10^{12}\, M_{\odot}}$, the ratios 
between different CO transitions and the LFs from these lines are poorly constrained and this work can serve as motivation to plan an experiment especially designed to measure CO emission from several rotational transitions and their redshift evolution.

\subsection{CO signal: intensity and power spectrum estimates}
\label{sec:CO_r}

We will now show theoretical estimates for the intensity and the power spectra of CO contamination using the LFs obtained with 
the two CO methods. The CO intensity can be obtained by integrating over the CO LF for the luminosity range available for each line. Following \cite{2012ApJ...745...49G} the CO intensity is given by:
\be
I^J_{\rm CO}(\nu)=\int^{L^{\rm J}_{\rm min}} _{L^{\rm J}_{\rm max}} dL \frac{dn}{dL} \frac{L^{\rm J}_{\rm CO}}{4 \pi D^2_L} y\left(z(\nu,J)\right) D^2_{\rm A},  
\ee
\\
where $dn/dL=\Phi(L)$.

When calculating the power spectrum from a CII map contaminated by CO emission, the corresponding CO power spectrum will be rescaled from 
the original value at the CO emission redshift, both in amplitude and in terms of the wavelength.
Following \cite{2014ApJ...785...72G} the contamination CO power spectra is given by: 
\ba
P_{\rm obs}(k_{\perp},k_{\parallel})&=& \left[ P^{\rm clus}_{\rm CO}(z_f,J,k_f) + P_{\rm CO}^{\rm shot}(z_f,J,k_f) \right] \nonumber \\
 &\times&  \left(\left[ \frac{\chi(z_{\rm s})}{\chi(z_f)}\right]^2 \left[\frac{y(z_{\rm s})}{y(z_f)}\right] \right),
\ea
where the clustering power spectra is given by:
\ba
P^{\rm clus}_{\rm CO}(z_f,J,k_f)&=& \bar{I}^2_{\rm f}(z_f) b^2_{\rm f}(z_f) P_{\delta\delta}(z_f,k_f).
\ea
The indexes $s$ or $f$ indicate whether we are referring to respectively the source
(CII) or the foreground (CO) redshifts, $\chi$ is the comoving distance, 
$|\vec{k_f}|\, =\, \left[ (r_s/r_f)^2 k^2_{\perp}\, +\, (y_s/y_f)^2 k^2_{\parallel} \right]^{1/2}$ is the three dimensional k vector at
the redshift of the foreground line, $P_{\delta\delta}$ is the matter power spectra and $b^{\rm J}_{\rm CO}$ is 
the bias between the CO$(J\to J-1)$ signal and dark matter.

The shot noise power spectra due to the discrete nature of galaxies is given by:
\be
P^{\rm shot}_{\rm CO}(z,J)=\int^{M_{\rm max}}_{M_{\rm min}} dM \frac{dn}{dM} \left[ \frac{L^J_{\rm CO}(M,z)}{4 \pi D^2_L}y(z,J)D^2_A \right]^2.
\ee

There are distortions in the observed power spectra in different directions due to redshift evolution of the 
signal. In theory these distortions could be used to differentiate between the signal and the foregrounds or to 
confirm if the foregrounds where effectively removed since in that case there should be no distortions observed (besides the known redshift-space distortions).
However, in practice this would require an experiment with an extremely high resolution and so in this study we 
will only consider the spherical average power spectra. The foreground lines will contaminate the spherical 
average CII power spectra at 
$|\vec{k}_s|=\left[k^2_{\parallel}+k^2_{\perp} \right] ^{1/2}$.

Since we assume that there is a correlation between CO luminosity and dark matter halo mass then the bias between the overall CO emission and the underlying dark matter 
density field can be estimated from the halo bias ($b(z,M)$) as:
\be
b^J_{\rm CO}=\frac{\int^{M_{\rm max}}_{M_{\rm min}} dM \frac{dn}{dM}L^J_{\rm CO}(M,z) b(z,M)}{\int^{M_{\rm max}}_{M_{\rm min}} dM  \frac{dn}{dM}L^J_{\rm CO}(M,z)} 
\ee
where $M_{\rm min}\, =\, M(L^{\rm J}_{\rm min})$, $M_{\rm max}\, =\, M(L^{\rm J}_{\rm max})$ and 
$L^{\rm J}_{\rm CO}(M,z)\, \propto\, M^{\alpha_{\rm CO}}$. The correct value for $\alpha_{\rm CO}$ changes with the 
galaxy mass, assuming the values from the galaxies in the Obreschkow simulation we have $\alpha_{\rm CO}\, =\, 1.5\pm\, 0.5$ 
for halos with $M{\rm <10^{12}\, M_{\odot}}$ and $\alpha_{\rm CO}\, <\, 1$ for higher mass halos. For the 
observationally based CO contamination power spectra we assume $\alpha_{\rm CO}\, =\, 1$ in 
our calculations.
The estimated contamination power spectra of CO emission in CII maps observed in the frequency range 
200 GHz to 300 GHz is shown in Figures \ref{fig:ps_3D_CO_Jb} and \ref{fig:ps_3D_CO_Jc} for respectively the Obreschkow CO model 
and the observational CO model.

\subsection{Contamination from atomic emission lines}
\label{sec:other_contamination_lines}

The $\rm [OI]\, 145\, \mu$m, the $\rm [NII]\, 121.9 \mu$m and the $\rm [NII]\, 205.2 \mu$m atomic emission lines are emitted from PDRs or 
from HII regions and so the luminosity of these lines is powered by stellar emission. Therefore it is expected to be 
correlated with the galaxies SFRs. 
The luminosity of these lines depends highly on the galaxies gas density and FUV flux \citep{1999ApJ...527..795K}. However, for a 
large number of galaxies, their luminosity densities scale with the FIR luminosity. We therefore used the observational ratios, presented 
in Table \ref{tab:line_emission_contaminants}, taken from
\citep{2011ApJ...728L...7G,2008ApJS..178..280B,2011ApJ...740L..29F,2013ApJ...765L..13Z}, to estimate the lines luminosities ($[x]/[{\rm FIR}]$ stands for the average fraction of FIR emission of each line).
The luminosity of these lines as a function of halo mass was then obtained using Equations \ref{eq:LIR_LFIR}, \ref{eq:L_IR_SFR}
and \ref{eq:SFR_param}.
The intensity of each line in the relevant range of 200 to 300 GHz, estimated using Equation \ref{eq:Int_CII} is shown in Table \ref{tab:line_emission_contaminants}.

\begin{table}[h]
\centering     
\caption{Average intensity of several emission lines in the observed frequency range 200 - 300 GHz in units of 
${\rm [Jy\ sr^{-1}]}$.}                   
\begin{tabular}{l  c  c  c  c}        
\hline\hline                 
$line$ & $[x]/[{\rm FIR}]$ & $z[200\, {\rm GHz}]$ & $z[300\, {\rm GHz}]$ & ${\rm Intensity}$ \\    
\hline                        
   ${\rm [OI]\, 145  \mu}$m  & $0.05\%$ &$9.3$ &  $5.9$ & $5.1$ \\
   ${\rm [NII]\, 122 \mu}$m  & $0.01\%$ & $11.3$  &  $7.2$ & $5.5$  \\
   ${\rm [NII]\, 205 \mu}$m  & $0.03\%$ & $6.3$  &  $3.9$ & $58$  \\
\hline                                  
\end{tabular}
\label{tab:line_emission_contaminants}     
\end{table}
The average contamination from these lines is considerably below the CII intensity.

\subsection{Contamination from continuum emission}
 The contamination from continuum emission can be estimated from the SFR and gas properties. 
The origins for the continuum emission considered here include: stellar continuum emission which escapes the galaxy, stellar emission 
reprocessed by the dust in the galaxy, free-free and free-bound continuum emission caused by interactions between free electrons and 
ions in the galaxies, and two photon emission originated during recombinations.
Since continuum radiation (with the exception of some bands in stellar continuum radiation) observed in the frequency range 200 - 300 GHz will be 
emitted in the infrared band, it will not be absorbed by any of the main hydrogen lines or by 
dust and so we can assume that this radiation is not affected during its path towards us. 

The intensity of contamination from all of the referred continuum sources of infrared emission is shown in Table 
\ref{tab:cont_contamination} and the detailed calculations are presented 
in the appendix.
\begin{table}[h]
\centering     
\caption{Intensity of continuum emission observed at frequencies of 300 GHz and 200 GHz in units of $[{\rm Jy\, sr^{-1}}]$}                    
\begin{tabular}{l  c c}        
\hline\hline                 
Source of emission & ${\rm I(300\, GHz)}$ & ${\rm I(200\, GHz)}$ \\    
\hline                        
   dust         & $3.0\times10^{5}$  & $2.0\times10^{5}$ \\
   stellar      & $4.1\times10^{-3}$ & $8.5\times10^{-1}$  \\
   free-free+free-bound    & $1.3\times10^{1}$ & $0.9\times10^{1}$ \\
   2-photon     & $3.4\times10^{-12}$ & $2.8\times10^{-12}$ \\
\hline                                  
\end{tabular}
\label{tab:cont_contamination}     
\end{table}

It is also expected that there is some contamination from the Milky Way which can be estimated from temperature maps 
of our galaxy for the relevant frequencies. Using temperature maps from Planck at frequencies ${\rm 143\, GHz}$, ${\rm 217\, GHz}$ and ${\rm 353\, GHz}$ 
we estimated that unless we were in the center of the Milky Way where the brightness temperature can reach 0.2 - 0.3 K, the 
average brightness temperature is well below $0.1\,$K which  corresponds to an observed intensity of ${\rm 2.44\,\times\, 10^{-32}\, Jy\,sr^{-1}}$ to 
${\rm 1.18\, \times\,10^{-32}\,Jy\,sr^{-1}}$ for 200 GHz and 300 GHz respectively.

The intensities in Table \ref{tab:cont_contamination} show that the continuum contamination is above the CII signal 
however continuum emission can be fitted and efficiently subtracted from the observational maps.

\section{Simulations of the observed signal}
\label{sec:Signal_generation}

The CII mock observational cone was made using the following steps:
\begin{enumerate}

\item A dark matter density field with a size of $L_{\rm box}\, =\, 634\, {\rm h^{-1}\, Mpc}$ and a number of cells of 
$N^3_{\rm box}\, =\, 1800^3$ was generated using the 
Simfast21 code \citep{2010MNRAS.406.2421S,2012arXiv1205.1493S}. 

\item The same code was used to generate dark matter halo catalogs from the previously 
generated density field using the excursion set formalism and by sampling the halos directly from the density field. These 
catalogs were made for redshifts 5.3 to 8.5 with 
a redshift step of 0.1 and a halo mass range of $10^8\, {\rm M_{\odot}}$ to $10^{15}\, {\rm M_{\odot}}$. At this point the halo
properties contained in the catalogs included only the halo mass and its position in a three dimensional box with the size 
and resolution of the density field.

\item  We randomly assigned astrophysical properties, such as SFR, from the 
\citet{2007MNRAS.375....2D} galaxies catalog, to the generated 
halos according only to the halos mass and redshift. 

\item We added CII luminosities to the halo properties using the halos SFR versus CII luminosity relations 
shown in section \ref{subsec: LCII_versus_SFR}, which resulted in four CII luminosity values for each halo, one for each 
of the ${ \bf m_1}$, ${ \bf m_2}$, ${ \bf m_3}$ and ${ \bf m_4}$ models.

\item In order to build the observational cones we made a box with 50 by $256^2$ cells which 
covers the frequency range 200 to 300 GHz and the $1.3\, {\rm deg}\, \times\, 1.3\, {\rm deg}$ field of view with steps of respectively 
$df_{\rm o}\, =$ 2 GHz and $d_{\rm ang}\, =\, 8\, \times\, 10^{-3}\, $deg. The angular 
coordinates correspond to positions in right ascension (RA) and declination (DECL), where 
the center of the box (the cone rotation axis) is at RA = 0 and DECL = 0.  
\label{item5}
\item We filled the box with the halos by assuming that the halos z direction corresponds to the direction of the line 
of sight and that moving in this direction is equivalent to moving in redshift. Since the size of the halo catalogs in the z 
direction is smaller than the comoving distance from redshift 5.3 to 
8.5 we piled the catalogs in order to cover all the needed distance range but we rotated the upper catalogs in order to not 
repeat structures in the line of sight direction.
The initial position of the halos was assumed to be at the comoving distance at which emitted CII 
photons are observed at a frequency of 300 GHz ($d_{\rm f300}$). The position ${\rm (x_i,y_i,z_i)}$ of the halos was 
assumed to be at a distance ($d_{\rm x}\, =\, {\rm x_i}\, \times\, d_{\rm r}\,-\, L_{\rm box}/2$, 
$d_{\rm y}\, =\, {\rm y_i} \times d_{\rm r}\, -\, L_{\rm box}/2,dist_{\rm z}\, =\, d_{\rm f300}\, +\, {\rm z_i}\, \times\, d_{\rm r}$), 
where $d_{\rm r}=L_{\rm box}/N_{\rm box}$, which corresponds to a comoving distance
$dist_c\, =\, ( d_x^2 + d_y^2 + d_z^2 )^{1/2}$ and to an angular position in right ascension and declination of respectively:

\be
{\rm RA} {\rm = arctan}\left( \frac{d_{\rm x}}{d_{\rm com}}\right),
\ee
\\
and

\be
{\rm DEC} {\rm = arctan}\left( \frac{d_{\rm y}}{d_{\rm com}}\right).
\ee
\\
Each comoving position was converted first to a redshift and then to an observed frequency using 
$\nu_{\rm obs}\, =\, \nu_{\rm CII}/(1+z)$. 
The halos were then distributed in the cone according to their angular position and observed frequency. For each halo 
catalog at a redshift $z$ we only used the halos with a redshift lower than ${z+dz}$.
\label{item6}
\item In each cell of the mock observing cone, the intensity was assumed to be given by a sum over the contribution from each galaxy as:

\be
I_{\rm CII}=\sum_i \frac{1}{df_{\rm o}}\frac{L_{\rm CII}(M,z)}{4\pi D^2_L}. 
\ee
\\

\end{enumerate}

\section{Simulations of the CO foreground contamination}
\label{sec:Foregrounds_generation}

The CO mock observational cones were made using the following steps: 

\begin{enumerate}

\item A dark matter density field with a size of $L_{\rm halos}^3\, =\, 296^3$ ${\rm h^{-3}\, Mpc^3}$ and a number of cells of 
$N_{\rm halos}\, =\, 1200^3$ was generated using the Simfast21 code. 

\item The same code was used to generate dark matter halo catalogs from the previously 
generated density field using the excursion set formalism and by sampling the halos directly from the density field. These 
catalogs were made for redshifts 0 to 2.5 with 
a redshift step of 0.1 and a halo mass range of $10^8\, {\rm M_{\odot}}$ to $10^{15}\, {\rm M_{\odot}}$. The halo
properties contained in the catalogs include only the halo mass and position in a three dimensional box with the size 
and resolution of the density field.

\item  We randomly assigned astrophysical properties, such as SFR, CO fluxes and visual absolute magnitudes, from the 
\citet{2007MNRAS.375....2D} and the \citet{2009ApJ...702.1321O} simulated galaxy catalogs, to the halos (e.g. we allowed some randomness 
in the astrophysical properties for halos with same mass and redshift, using distributions from the SAX simulation).

\item We calculated the CO luminosity for each halo from its CO flux.

\item We repeated steps \ref{item5} and \ref{item6} of the CII signal generation assuming that 
the position z=0 of the halos corresponded to 
a comoving distance of zero and that the redshift could be converted to an observed frequency using 
$\nu_{\rm obs}\, =\, \nu_{\rm CO}^{\rm J}/(1+z)$.

\item For each CO transition we built mock observing cones with intensities estimated as in the CII 
case and then we added the cones to obtain the total CO intensity.

\item In order to simulate the effect of the masking technique we made a mock observing cone with only the galaxies with 
CO fluxes above a given threshold. The pixels with at least one galaxy correspond to pixels that should be masked in 
order to decrease the CO contamination in CII observational maps and so we put to zero the corresponding pixels in the initial 
CO box and in the CII box. We also used the same technique with a limit in magnitudes in the AB system K filter ($m_{\rm K}$) 
instead of a limit in CO flux.

\end{enumerate}

A slice of a mock observational CII cone is shown in Figure \ref{map}. This figure
shows that CII emission is not randomly distributed but that it follows the underlying density fluctuations.
\begin{figure}[htbp]
\hspace{-4mm}
\includegraphics[scale = 0.26]{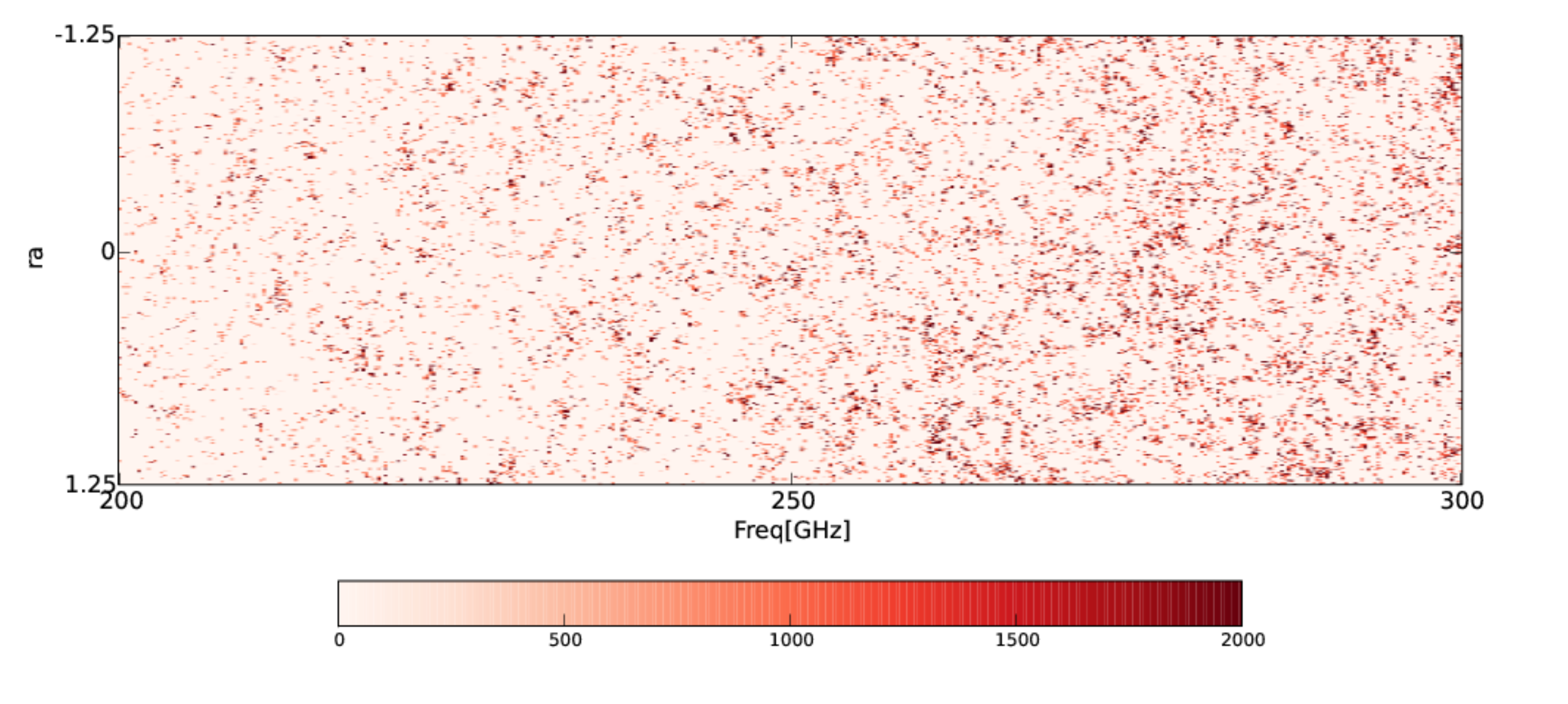}
\caption{Slice of a CII mock observing cone with 2.5 square degrees from frequencies 200 GHz to 300 GHz.}
\label{map}
\end{figure}

One of the advantages of simulating mock observational cones is that we can directly add the signal and its contaminants 
to obtain a more realistic version of an observational intensity map.
This is useful because it gives us better predictions of what an observational experiment will actually measure and how 
to relate the  observed signal to the intrinsic signal which is where the scientific information really lies. The analysis of the information contained in these cones is mainly made using the power spectra of the target emission 
line and so we used slices in frequency (these slices correspond to the signal emission around a given redshift) from these cones to construct intensity maps in Cartesian coordinates, from which we calculated the signal power spectra. 
With this method we directly mapped the contaminants intensity spatial fluctuations into Cartesian coordinates at the signal 
redshift. This allowed us to directly obtain a contamination power spectra which is essential to determine the real degree of foreground contamination and to plan ways to clean observational maps.

The CII intensities obtained with these observing cones is shown in Table \ref{tab:CII} together with the overall CO 
intensity from transitions (2-1) to (6-5) at the same observed frequencies. The results show that CO contamination will 
dominate observations especially for low frequencies. 
\begin{table}[h]
\centering      
\caption{Intensity of CII emission from galaxies as a function of redshift calculated using the SFR, The medium, maximum 
and minimum values of the CII intensity correspond to the LCII versus SFR parameterizations ${\bf m_2}$, ${\bf m_1}$ 
and ${\bf m_4}$ respectively. Also shown are the CO intensities estimated using the Obreschkow CO model. The intensities 
have units of ${\rm [Jy\, sr^{-1}]}$.}                  
\begin{tabular}{l  c  c  c  c}        
\hline\hline                 
$z_{\rm CII}$ & $\bar{I}_{\rm CII}$ & $I_{\rm CII}^{\rm max}$ & $I_{\rm CII}^{\rm min}$ & $I_{\rm CO}$\\    
\hline                        
   8.5  & $47.0\times10^{\rm 1}$  & $9.50\times10^{\rm 1}$ & $1.80\times10^{\rm 1}$  & $1.05\times10^{\rm 3}$ \\
   7.5  & $1.00\times10^{\rm 2}$  & $2.00\times10^{\rm 2}$ & $3.70\times10^{\rm 1}$  & $1.25\times10^{\rm 3}$ \\
   6.5  & $1.90\times10^{\rm 2}$  & $3.50\times10^{\rm 2}$ & $7.50\times10^{\rm 2}$  & $1.12\times10^{\rm 3}$ \\
   5.5  & $3.36\times10^{\rm 2}$  & $6.00\times10^{\rm 2}$ & $1.33\times10^{\rm 2}$  & $1.06\times10^{\rm 3}$ \\
\hline                                  
\end{tabular}

\label{tab:CII}     
\end{table}

\section{Instrument parameters}

The characteristics of an experiment able to measure the CII intensity and spatial fluctuations will now be briefly discussed.

\begin{table}[t]
\begin{center}
\caption{Parameters for a CII experiment }
\vspace{4mm}
\label{tab:Cospec}
\begin{tabular}{l | c | c }
\hline\hline
Instrument & CII-Stage I & CII-Stage II \\
\hline
Dish size   ($\rm m$) & $10$ & $10$\\
Survey area $A_{\rm s}$ ($\rm arcmin^2$) & $78 \times 0.5$ & $600\times 600$\\
Instantaneous FOV ($\rm arcmin^2$) & $13.6\times0.5$ & $25.6\times0.4$ \\
Freq. range (GHz) & 200 - 300 & 200 - 300 \\
Frequency resolution (GHz)& 2 & 0.4 \\
Number of Spectrometers & 32 & 64 \\
Total number of bolometers  & 1600 & 16000  \\
On-sky integration time (hr) & 1000 & 2000 \\
NEFD on sky (${\rm mJy\, \sqrt{sec}}$) & 65 & 5  \\
\hline
\end{tabular}
\label{tab:Cospec}
\end{center}
\end{table}

We propose to use one of two similar setups, the first one (CII-Stage I) is appropriate for optimistic CII models (models with a high CII 
luminosity density) and the second 
one CII-Stage II has the minimum requirements to insure a CII power spectra detection in the case of a more pessimistic CII model.
The choice of a setup for the CII experiment is mainly dependent on the evolution of the CII luminosity for high redshifts and 
so it can be updated when more high redshift CII observations are available.

The basic experiment proposed here (CII-Stage I) consists in using one stack of independent single beam, single polarization spectrometers, 
one stack for each polarization. Each of these spectrometers would contain several bolometers and each of the 
stacks would cover a line on the sky via a polarizing grid. The second stage experimental setup (CII-Stage II) is similar to the first 
one but covers a much larger area with a narrower spectra. The details of the proposed experimental setups are shown 
in Table \ref{tab:Cospec}. The angular resolution of the experiments is ${\rm (0.5\, arcmin)^2}$ for CII-Stage I 
and ${\rm (0.4\, arcmin)^2}$ for CII-Stage II, respectively.

\section{Cleaning contamination from lower redshift emission lines}
\label{sec:Foreg_removal}
\subsection{Pixel masking}

As was shown in previous sections, the main line contaminants for the planned observations are CO rotation lines from low redshifts. 
Since the contamination from CO emission lines in the CII power spectra is high, we made CO flux cuts to study which galaxies 
are dominating the contamination and if they can be removed from the observational data by masking the pixels with the stronger 
contaminants. 

\begin{figure}[hbp]
\hspace{-6mm}
\includegraphics[scale = 0.48]{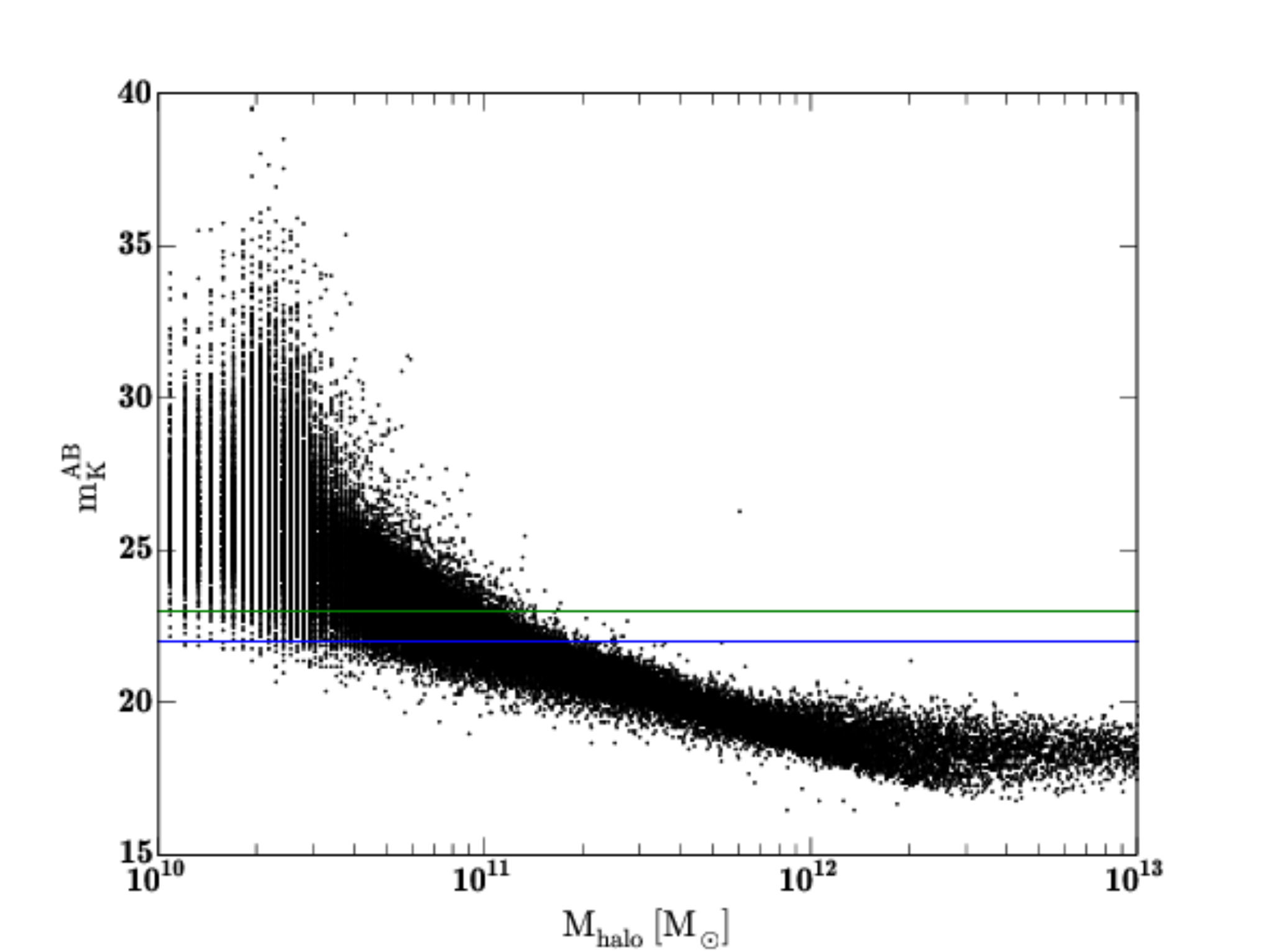} 
\caption{AB magnitude in the K filter versus halo mass relation at redshift z=0.06. The relation showed has a small 
evolution with redshift. The horizontal lines show what galaxies are being masked according to the ${\rm m_{\rm K}}$ cuts showed 
in figure \ref{fig:Ps_Jy_z65}.}
\label{fig:Mk_mass}
\end{figure}

\begin{figure*}
\centerline{
\includegraphics[scale = 0.32]{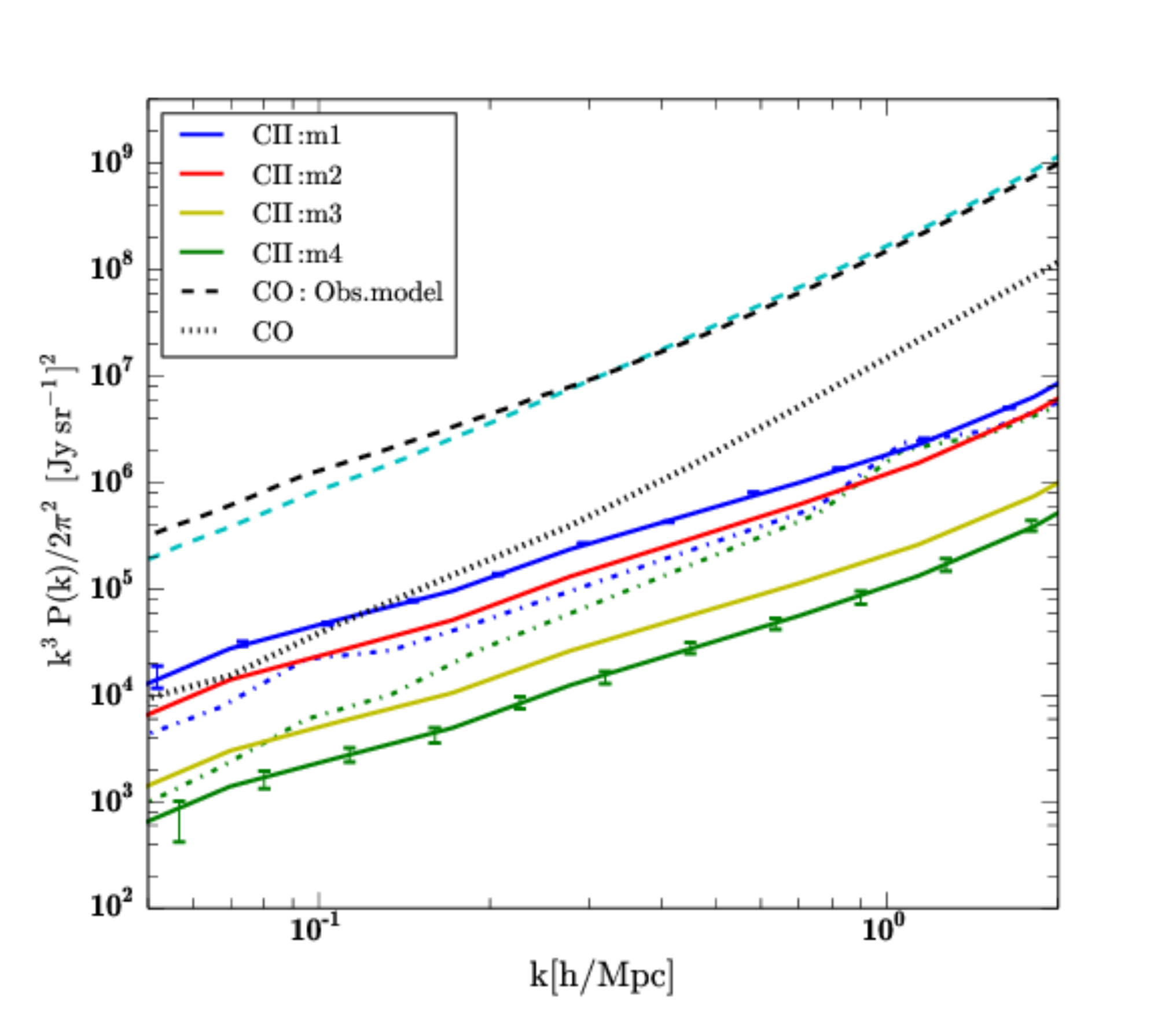}
\hspace{-8 mm}
\includegraphics[scale = 0.32]{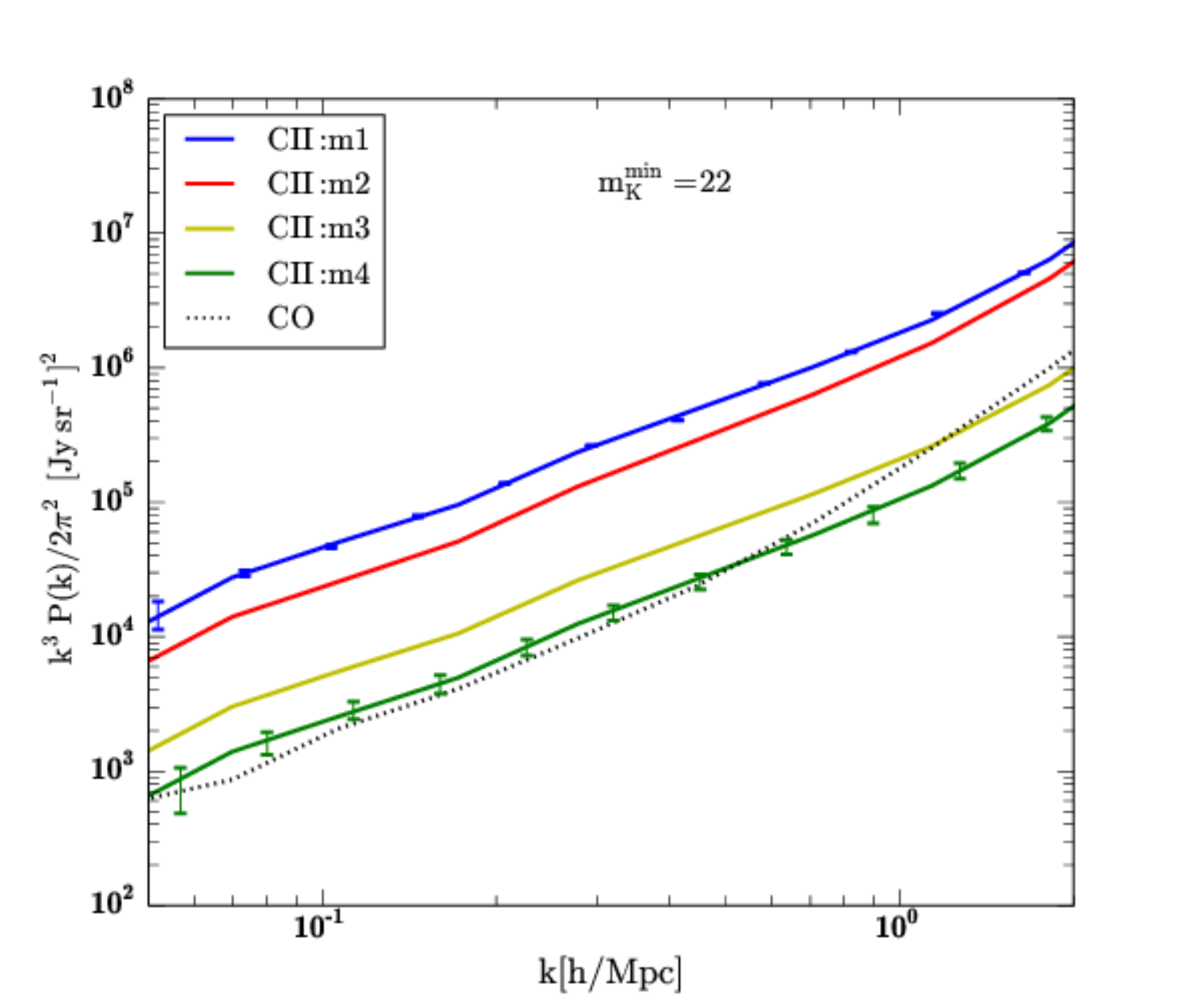}
\hspace{-8 mm}
\includegraphics[scale = 0.32]{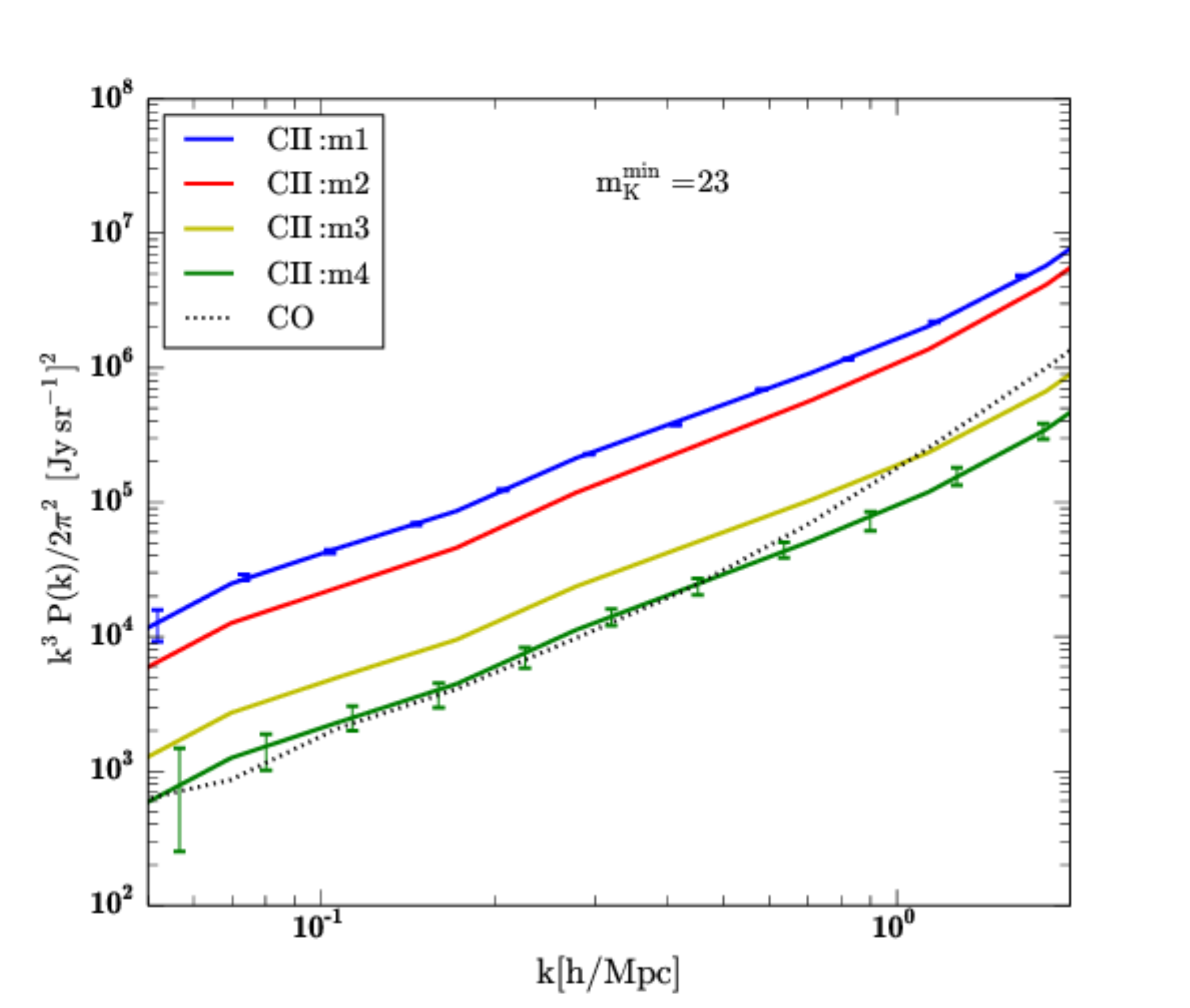}
}

\centerline{
\includegraphics[scale = 0.32]{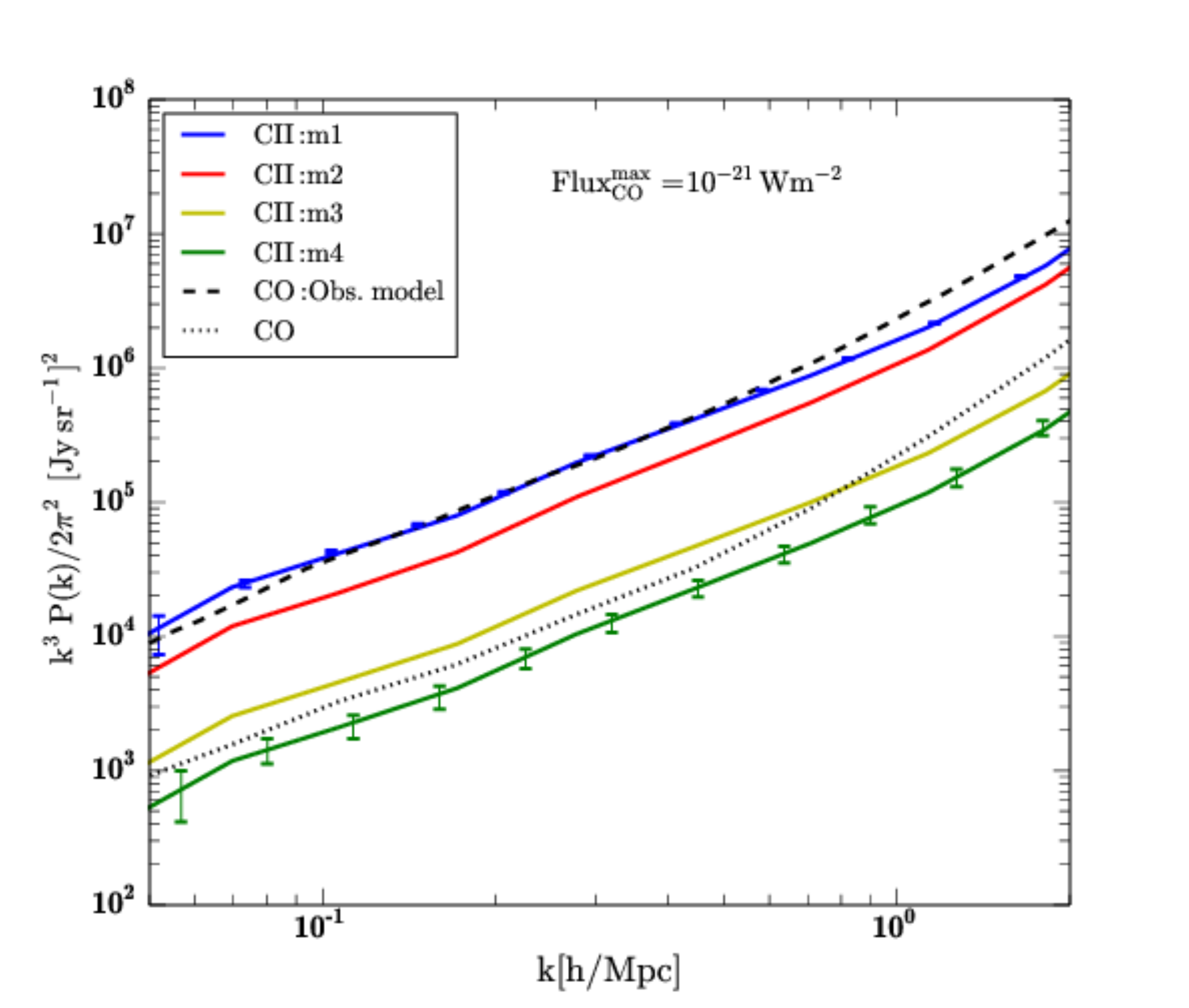}
\hspace{-8 mm}
\includegraphics[scale = 0.32]{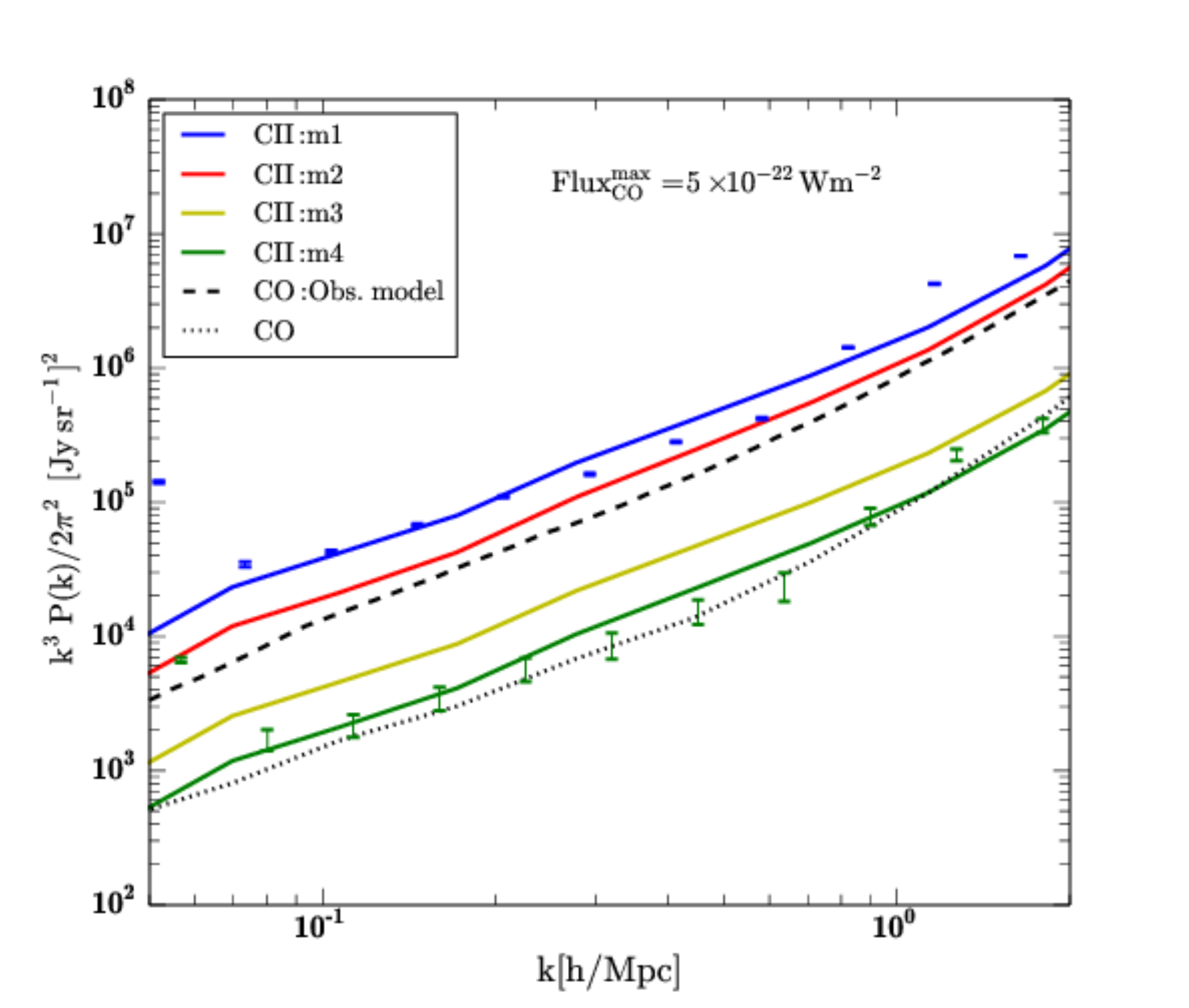}
\hspace{-8 mm}
\includegraphics[scale = 0.32]{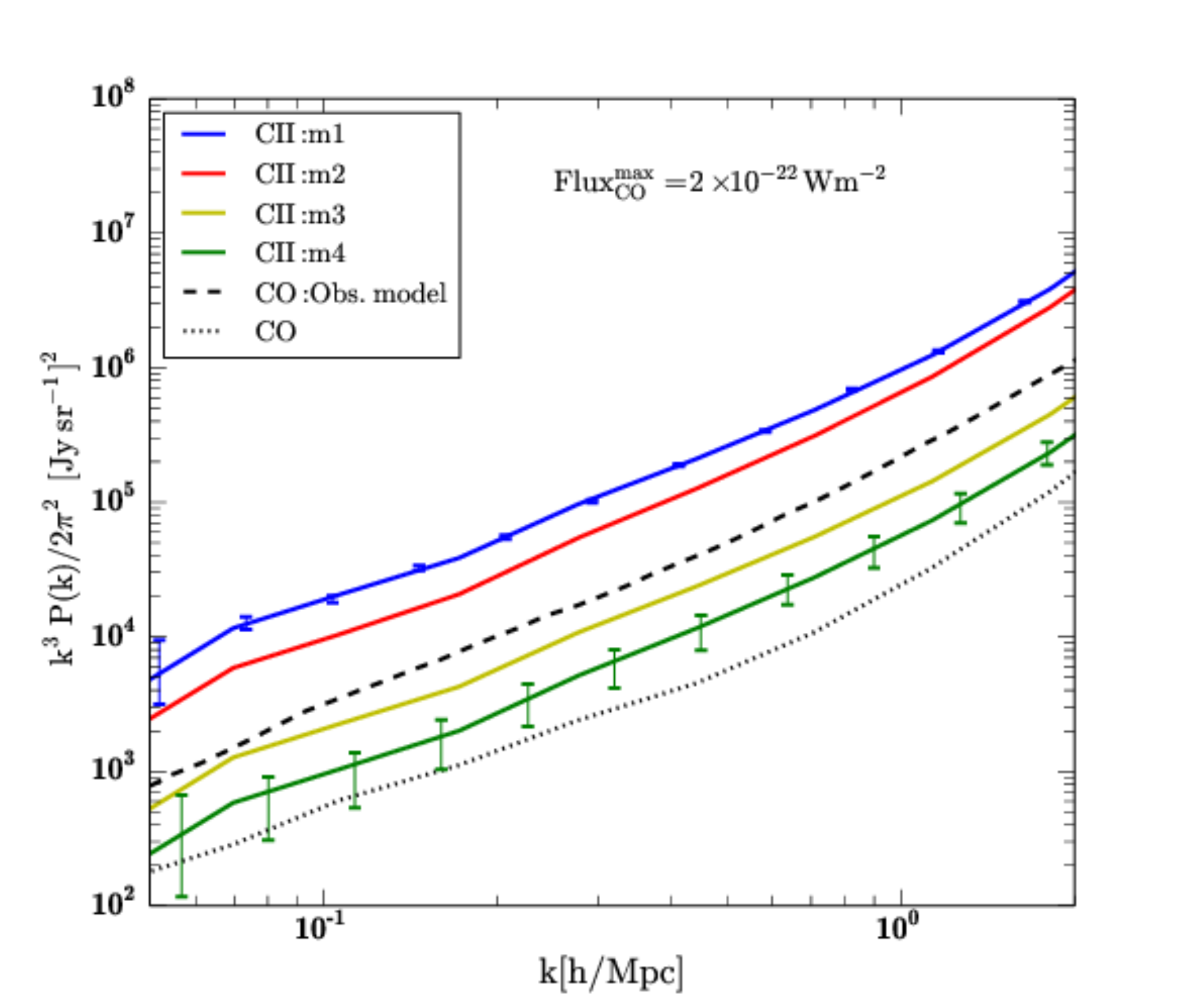}
}
\caption{\label{fig:Ps_Jy_z65} Power spectra of CII emission for four parameterizations of LCII vs SFR (solid lines) and 
power spectra of CO contamination (dotted lines) observed in a frequency range of 37GHz centered at 
${\rm F_{obs}\, \approx\, 250\, GHz}$. The cyan dashed dotted line corresponds to an upper limit for CII emission from ionized regions. Upper left panel corresponds 
to the total CII power spectra and CO contamination power spectra. 
The middle and right upper panels assume only CO sources with AB relative magnitudes in the K band above ${\rm m_{K}\, =\,22}$ 
and ${\rm m_{K}\, =\, 23}$ respectively.
Lower panels assume only CO sources with fluxes below ${\rm 10^{-21}\, Wm^{-2}}$ (left panel), ${\rm 5\, \times 10^{-22}\, Wm^{-2}}$ 
(middle panel) and ${\rm 2\, \times10^{-22}\, Wm^{-2}}$ (right panel). Also shown in the upper left and in the lower panels are 
dashed lines of the theoretical CO emission, with the same flux cuts, estimated using the CO observationally based model. The 
error bars in the CII models {\bf m1} and {\bf m4} are based in the experimental setup of the CII-StageII experiment
described in Table \ref{tab:Cospec}. Also shown in dashed dotted lines in the top left panel are the errors in the CII models 
{\bf m1} and {\bf m4} based in the experimental setup of the CII-StageI experiment described in Table \ref{tab:Cospec}.}
\end{figure*}

Given that detecting galaxies with low CO fluxes 
can be very challenging, we also consider using a CO tracer easier to detect such as the SFR or the relative magnitude in a 
broad band filter such as the K filter (measured as magnitudes in the AB system in the K filter ${\rm m_{\rm K}}$) which is centered at 
2190 nm and covers around 390 nm. The 
SFR can only be used as a CO tracer in star forming galaxies. Since there is also an intense CO emission in galaxies 
powered by active galactic nuclei, if we want to use SFR or infrared emission as a CO tracer we should use an additional 
tracer like observations in the visible band to target the active galactic nuclei. 

In Figure \ref{fig:Mk_mass} we can see that galaxies with a high CO flux also have relatively low magnitudes in the 
K band. Thus we estimated the ${\rm m_{\rm K}}$ cut necessary to reduce the power spectra of CO contamination.

We show in the bottom panels of Figure \ref{fig:Ps_Jy_z65} that for the more optimistic CII models the 
power spectra of CO contamination can be efficiently reduced by removing from the observational maps contamination by galaxies with CO fluxes 
in one of the CO rotation lines higher than ${\rm 5\, \times\, 10^{-22}\, Wm^{-2}}$ and that this can be done by masking less than 10$\%$ of 
the pixels for an experiment with a setup similar to the CII-Stage II experimental setup. In alternative the top panels of this figure show that masking in ${\rm m_{\rm K}}$ magnitudes is
also possible and the necessary masking would require a cut of ${\rm m_{\rm K}\ =\ }$22 in order to sufficiently decrease the power 
spectra of CO  contamination predicted for a CII model like ${\rm \bf m_2}$. 

For CII models which yield lower intensities the ${\rm m_{\rm K}}$ cut 
would have to be of ${\rm m_{\rm K}\, =\, }$23 or even higher which would make it impossible to apply the masking technique. 
The CO masking can be done with cuts in quantities like the CO flux, SFR, IR luminosity, magnitude in a given band or a 
combination of probes depending of the CO tracer experiments available. The masking cuts considered in this study 
are presented in Table \ref{tab:CO_masking} following the CII-Stage I or CII-Stage II experimental setups and assuming the Obreschkow CO model. The observational 
model would require masking CO galaxies till a flux cut of ${\rm f_{CO}\, =\, 2 \times 10^{-22}\, W\, m^{-2}}$ which corresponds to a masking percentage 
of $10\%$ or $21\%$ for the CII-Stage II and CII-Stage I experimental setups respectively.

\begin{table}[h]
\centering     

\caption{ Masking percentages for an observation in the frequency range from 200 to 300 GHz}                    
\begin{tabular}{l  c c}       
\hline\hline                 
Flux/${\rm m_{\rm K}}$ cuts &   CII-Stage I   & CII-Stage II \\ 
                &    Masking $\%$   &Masking $\%$ \\
\hline                       
   ${\rm f_{\rm CO}>1\times10^{-21}}[{\rm W\, m^{-2}}]$   &$7.70$& $1.97$ \\
   ${\rm f_{\rm CO}>5\times10^{-22}}[{\rm W\, m^{-2}}]$ & $12.99$& $3.39$   \\
   ${\rm f_{\rm CO}>2\times10^{-22}}[{\rm W\, m^{-2}}]$  &$23.31$ & $6.40$    \\
   ${\rm m_{\rm K}<22}$                   & $6.23$ & $1.58$   \\
   ${\rm m_{\rm K}<23}$                   & $13.76$ & $3.60$   \\
\hline                                  
\end{tabular}

\label{tab:CO_masking}     
\end{table}

If we are able to measure CO luminosities of some galaxies to a high precision it will be possible 
to remove their intensity from each pixel instead of masking the pixel completely. This would 
reduce the masking percentage. However the number of galaxies
which we can observe with the necessary precision to remove their contamination from observations accurately should be rather small. Also, in order to do this, the 
intensity of the galaxy would have to be above the ``noise" in each of the pixels.

\subsection{Cross correlating foregrounds}

In this section we discuss the possibility to use cross correlation as a method to help removing CO foregrounds from 
CII maps and as a way to probe the degree of CO contamination remaining in CII maps after the masking technique has 
been applied.

\subsubsection{Cross correlation with galaxy surveys}

First, we consider cross correlating a CO line with the number density of galaxies.

As is shown in Figure 
\ref{fig:cross_map} the intensity of CO emission in the CO(5-4) line is strongly correlated with the 
galaxies number density at the same redshift since they both trace the underlying dark matter 
density fluctuations. 
\begin{figure}[htbp]
\center{
\vspace{0.1cm}
\includegraphics[scale = 0.49]{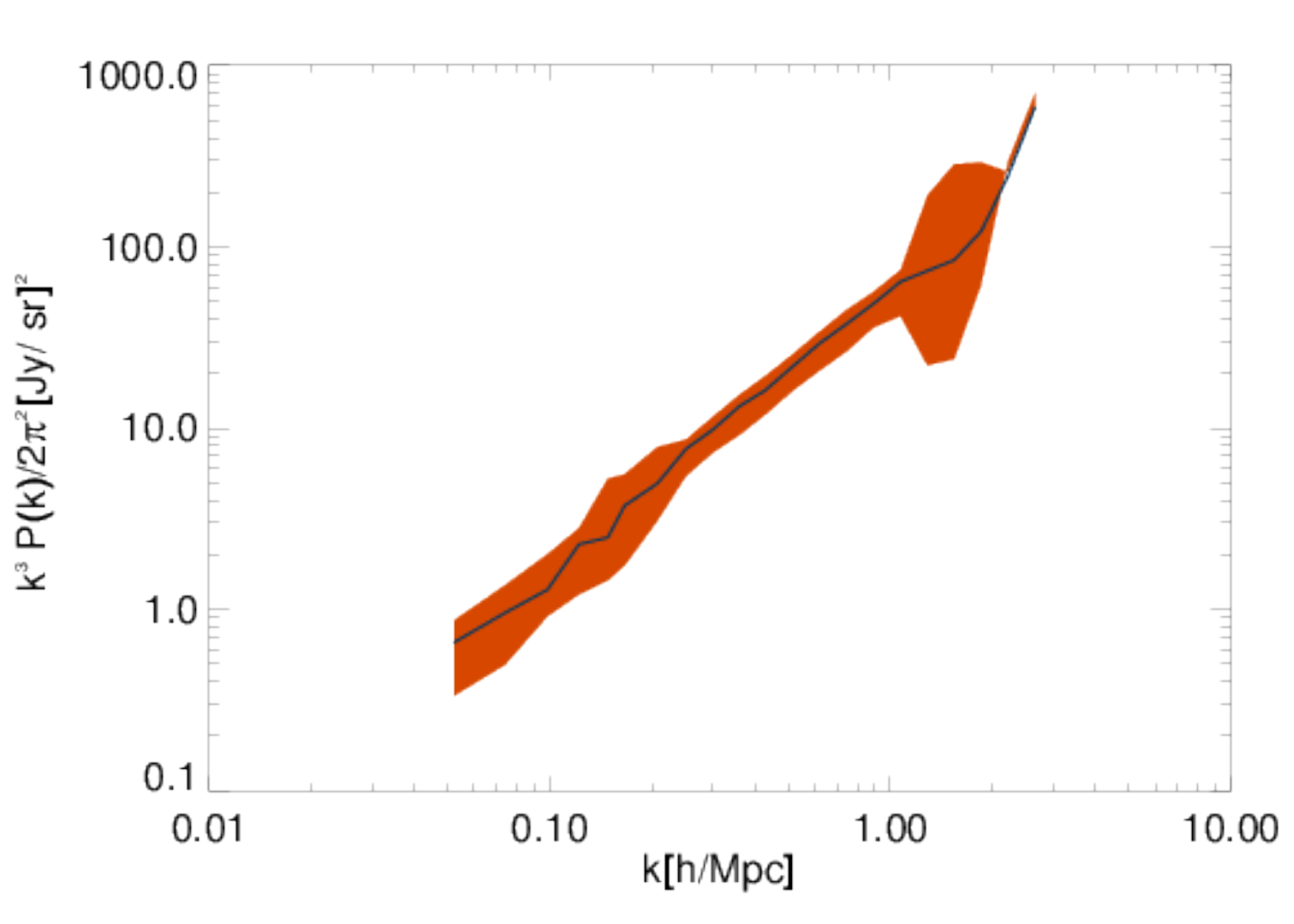}

\caption{\label{fig:cross_map} Cross correlation power spectra between number of galaxies and intensity of the CO(5-4) 
line at redshift z=1.4 obtained from our simulations. The 1$\sigma$ uncertainty shown in orange was obtained by cross correlating these two quantities 
in different regions of the space.}
}
\end{figure}
Here we consider the case that the number density of galaxies is independently measured with a galaxy survey. The number density of 
galaxies at a redshift z=1.4 can be cross-correlated with an observational intensity map of the CO(5-4) line centered 
at the same redshift (obtained 
from the 200 - 300 GHz observing cone) and the result will be proportional to the intensity fluctuations of the CO(5-4) line
even if the intensity map also contains CII and other CO lines.
This can be done for several foreground lines and redshifts to probe the degree of contamination  by these lines.\\

\subsubsection{Cross correlation between two CO lines}

As can be observed in Table \ref{tab:CO}, in some cases there are two CO lines originated from the same redshift contaminating 
the observational maps at two different frequencies. 
For example, the CO(3-2) and the CO(4-3) emitted at a redshift of $z\, =\, 0.6$ 
will be observed at frequencies of 288.2 GHz and 216.1 GHz respectively, and so they will contaminate CII intensity maps at 
redshifts 7.8 and 5.6. 

\begin{figure}[htbp]
\hspace{-8mm}
\includegraphics[scale = 0.41]{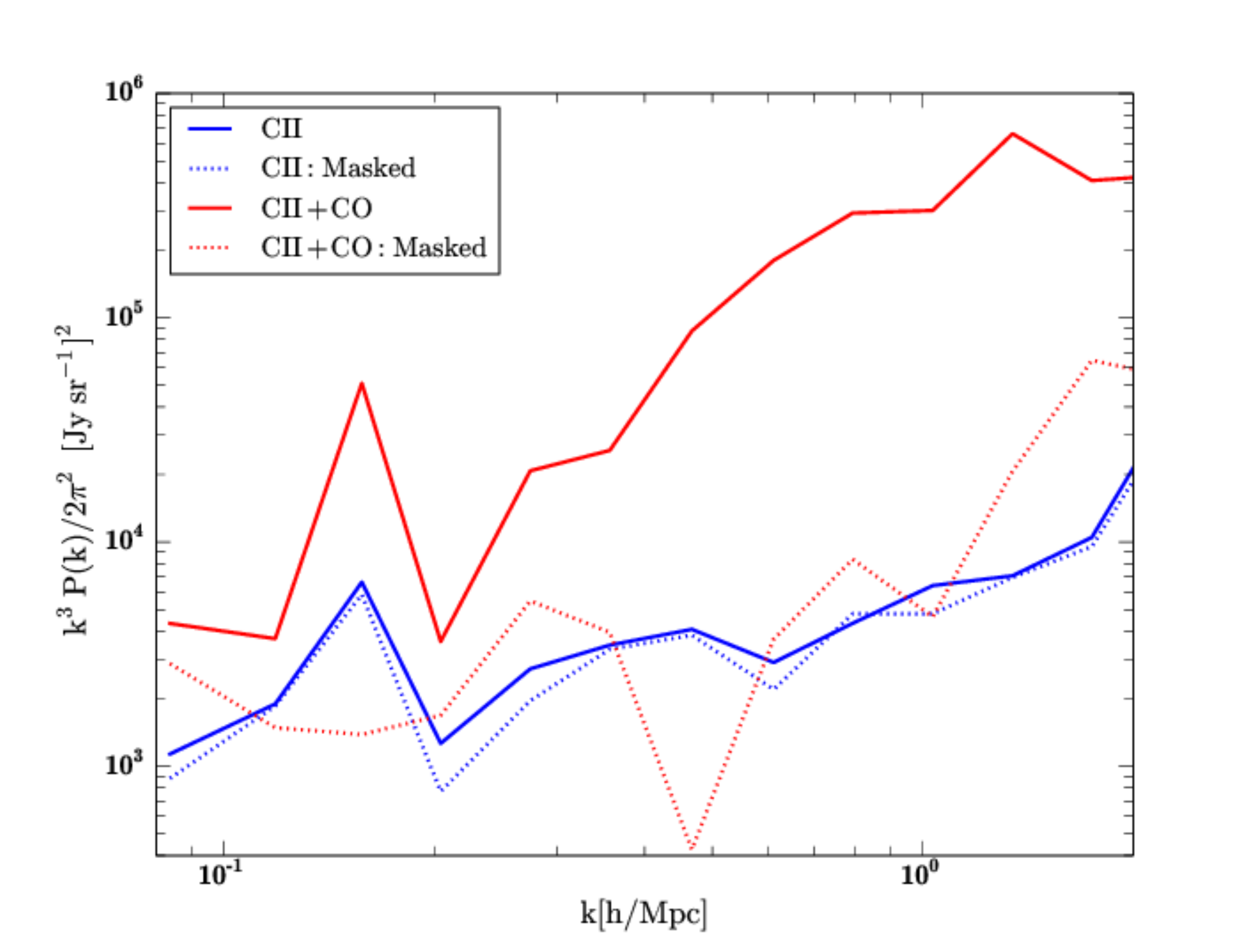} 
\caption{Cross correlation power spectra between observational maps with 70 Mpc centered at frequencies 288.2 GHz and 216.1 GHz 
which correspond to CII emission from redshift 7.8 and 5.6. The blue thin lines shows the cross correlation obtained from
maps with only CII emission. The red thick lines shows the cross correlation obtained from maps with CII plus 
CO. Solid line denote the full signal while dotted lines denote the signals masked till a CO flux of 
$2\times10^{-22}\, {\rm W\, m^{-2}}$. The masking was done assuming the CII-Stage II experimental setup.}
\label{fig:ps_cross_CO_CII_slices}
\end{figure}

As is shown in Figure \ref{fig:ps_cross_CO_CII_slices} the cross correlation between observational maps
with CII plus CO will be  stronger than maps with just CII and so by cross correlating intensity maps before and 
after masking, we can confirm if the cleaning procedure was successful.
Also, the existence of contamination from two lines emitted from the same redshift can in principle be  implemented in algorithms 
to help removing the CO contamination, although that task is out of the objectives of this study.

\section{Cross correlation between the HI and the CII lines}
Both fluctuation in HI and in CII intensity maps are correlated with fluctuations in the underlying density field and so 
the {spatial distribution of emission in these two lines is correlated. Therefore, the cross correlation power spectra of the 
the two lines gives a measure of their  intensities}.

Since CII is emitted from galaxies and HI is emitted from the IGM these two quantities are mostly negatively correlated 
at large scales. At small scales the correlation between CII and HI emission should be positive since they are both biased 
in overdense regions. However, we find no correlation in our simulations, which is probably due to the low intensity of 21 cm emission at theses scales.

\begin{figure}[htp]
\begin{center} 
\includegraphics[scale = 0.47]{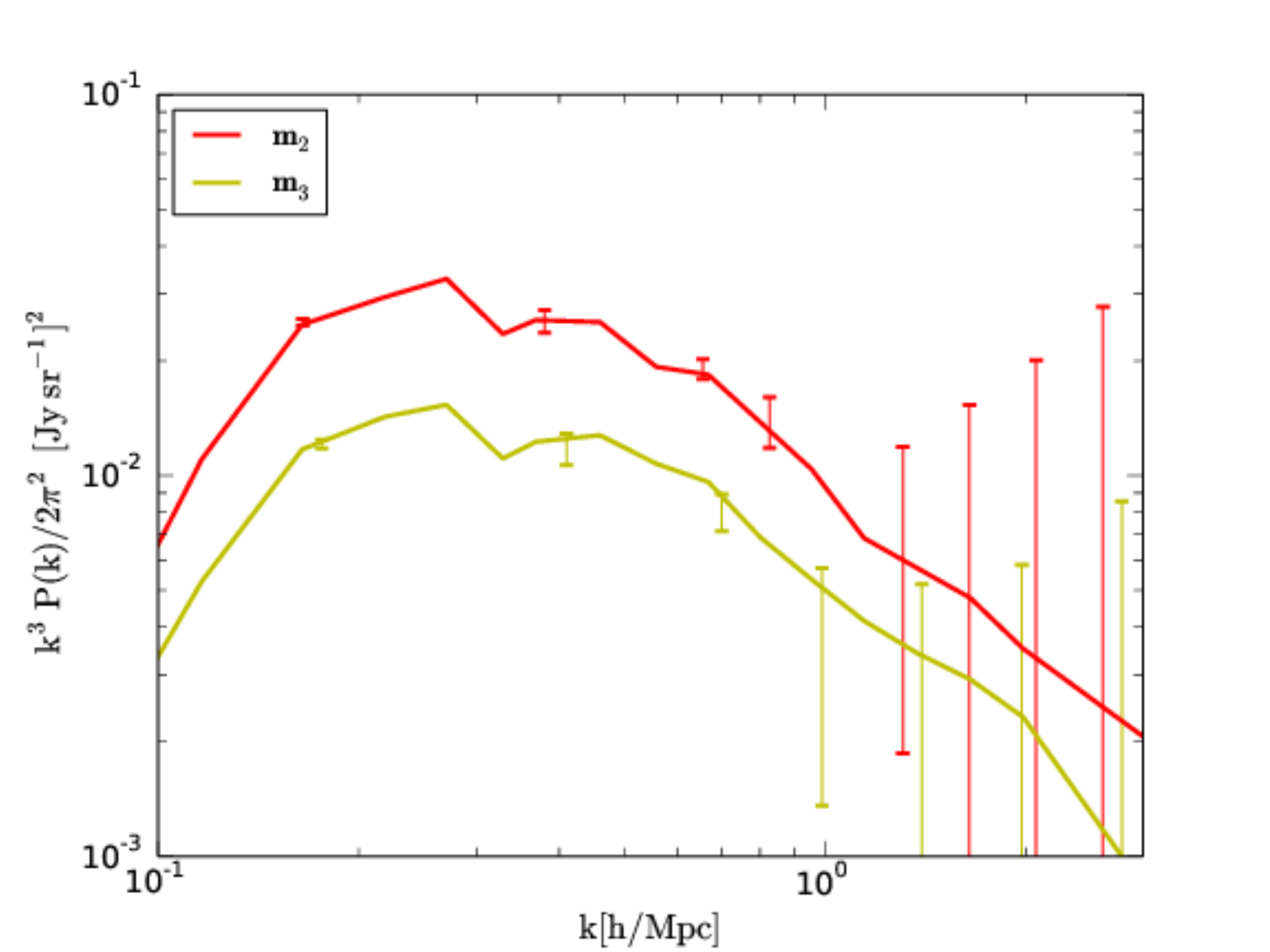} 
\caption{\label{fig:cross_HI_CII} Absolute value of the correlation power spectra between CII emission (models ${\rm \bf m_2}$ and
${\rm \bf m_3}$) and the
neutral hydrogen 21 cm line at redshift z=7.0. The error bars were estimated assuming the CII-Stage II experimental setup. Note 
that these two lines are negatively correlated at the shown scales.}
\end{center}
\end{figure}

In Figure \ref{fig:cross_HI_CII} we show the cross power spectra between HI emission and two models for CII emission.
The error bars shown in this figure were obtained with the HI 21 cm line experiment described in Table \ref{tab:cospec} and 
with the CII-Stage II instrument.

\begin{table}[htb]
\begin{center}
\caption{Parameters for SKA1-low }
\vspace{4mm}
\label{tab:cospec}
\begin{tabular}{l | c | c}
\hline\hline
& SKA1-low(z=7) &unit\\

\hline
station diameter $D_{\rm stat}$  & 35 &m\\
survey area $A_{\rm s}$ & 6.55&$\rm deg^2$\\
FoV per station & 6.55 &$\rm deg^2$\\
effective area per stat. $A_e$ & 355.04 & $\rm m^2$\\
freq. resolution $d\nu$ & 3.9 & kHz\\
bandwidth ($z\, =\, 8\, \pm\, 0.5$) BW & 18 &  MHz \\
tot. int. time $t_{\rm int}$ & 200& hr\\ 
min. baseline $D_{\rm min}$ & 35 & m\\ 
max. baseline $D_{\rm max}$ & 1 & km\\ 
collecting area $A_{\rm coll}$ & $307466$  &$\rm m^2$\\
$\rm uv_{\rm min}$ & $21$  & \\
$\rm uv_{\rm max}$ & $596$  & \\
$\rm T_{\rm sys}$ & 291 &  K\\ 
effective num. stat. & 866 &  \\
\hline
\end{tabular}
\end{center}
\end{table}

\section{Conclusions}

In this paper we consider the possibility of applying the intensity mapping technique to the CII line at high redshifts in 
order to probe the EoR and galaxy properties in the early Universe. The ionized carbon CII 158$\mu$ m line is one
of the strongest emission lines in the spectra of star forming galaxies and so observing this
line is one of the few possible ways to study very distant galaxies. Given the uncertainty in CII emission from high redshift galaxies 
we took into consideration four models for CII emission which cover the uncertainty in the relation between CII emission and SFR. We concluded 
that intensity mapping of the CII line during the end of the EoR is in the reach of today's technology.

The intensity of the CII signal from $z\, \sim\, 8.5$ to $z\, \sim\, 5.5$ is likely to be between $10^2$ and $10^3{\rm Jy\, sr^{-1}}$ although 
higher values would be possible if the SFRD is higher than the current predictions.
In the local universe, CII emission from star-forming galaxies is a good probe of their SFR and so intensity mapping of 
this line should provide good constraints on the SFRD at high redshifts, even if the constant of proportionality between CII luminosity and SFR 
evolves with redshift. Although a reasonable dispersion in the CII emission versus SFR relation is expected, in intensity mapping studies, we are averaging the relation over 
thousands galaxies in each pixel, so that the total CII emission should be averaged by the SFRD. The CII intensity should be dominated by galaxies 
with luminosities below the threshold of galaxy surveys and so even if CII emission in bright galaxies is not a perfect tracer for star formation, 
it should provide good constraints in the SFRD.
The CII line is also dependent in the ISM metallicity and although CII intensity maps cannot give strong constraints to this quantity they will 
provide a lower limit which will be on its own an improvement over current constraints of the gas metallicity at high redshifts. Note that 
redshift evolution of the metallicity can tell us about the characteristics of POP II and POP III stars 
which is also particularly important for Reionization studies.

Emission from CO rotation lines is going to be the main contaminant in CII observations and although the 
CO and CII intensities have a large uncertainty it is reasonably confirmed that some of the CO signal has to be removed from 
observations in order to recover the correct CII fluctuations.  
We estimated the CO intensity using two independent methods, one based in detailed simulations of gas conditions in galaxies 
and physical relations between CO transitions and other which uses only observational quantities and observational based relations
between these quantities. 
Both these methods predict similar CO intensities. 
The current constraints in CO and CII emission indicate that the CO power spectra will be up to one order of magnitude higher 
than the CII power spectra. However we showed that the CO signal can be reduced at least as much, by masking the pixels contaminated by the galaxies with the brighter CO emission.

We described an experiment which is within reach of current technology and is able to measure the CII power spectra with 
enough resolution so that we can mask most of the CO contamination without erasing the CII signal. 
In order to identify the most luminous CO galaxies we propose to use a galaxy survey able to measure CO luminosities or a 
more modest survey able to detect the galaxies AB magnitudes in the K band, since this is a good tracer of CO luminosity.
A galaxy survey able to measure CO luminosities of several transitions till a redshift of at least 2.5 would also provide the 
first LFs for CO transitions higher than the first CO rotational transition which would by it self be a valuable 
contribution to the study of gas conditions of galaxies.

If the CO contamination is too high and the masking technique is not enough to successfully clean the 
images or in order to confirm if the contamination was well removed, we can use cross correlations between different 
CO lines to estimate the intensity of their contamination.
Even in the worst case scenario where the overall CO emission is a few times higher than what we have considered, we still would be able to remove CO to at 
least detect the CII signal with the proposed CII-Stage II experimental setup. Moreover cross correlation between intensity maps of CII and other lines from 
the same redshift will not suffer from line contamination.

Finally the CII line and the 21 cm line are expected to be strongly anticorrelated. By cross correlating CII and 21 cm maps, 
we will obtain a statistical estimate of the intensity of these signals independent of most foregrounds which can be a valuable 
asset in constraining Reionization.

\begin{acknowledgments}
This work was supported by FCT-Portugal with the grant SFRH/BD/51373/2011 for MBS and  
under grant PTDC/FIS-AST/2194/2012 for MBS and MGS.\\ 
MGS was also supported by the South African Square Kilometre Array Project and the South African National Research Foundation.\\
AC and YG acknowledge support from NSF CAREER AST-0645427 and AST-1313319 at UCI and also from the Keck Institute for Space Studies (KISS) subcontract for intensity mapping studies.\\
MBS was also a long Visiting Student at UCI, supported by NSF CAREER AST-0645427 and AST-1313319 and she thanks the Department of Physics and Astronomy at UCI for hospitality during her stay.\\
We thank Jamie Bock, Matt Bradford and the TIME team for useful discussions.

\end{acknowledgments}

\bibliographystyle{apj}
\bibliography{apj-jour,CII_IM}

\appendix

In this appendix we summarize the key steps necessary to obtain the continuum foregrounds which will contaminate CII intensity maps in the 
frequency range 200 - 300 GHz. This study includes contamination by stellar emission, dust emission, free-free and free-bound emission and finally 
two photon emission.\\

\subsubsection{Stellar emission}
\label{sec:Stellar_cont_emission}

The stellar luminosity at frequency $\nu$ is approximately given by the emissivity of a black body ($B_{\nu}$) integrated 
over the solid angle and the area of the stellar surface ($ 4\pi R^2_{\ast}$):

\be
L^{\star}_{\nu}=\pi\, 4\pi\, R^2_{\ast}\, B_{\nu}\,(T^{\rm eff}_{\ast}). 
\label{eq:Lstar}
\ee
\\
For estimating the stellar radius ($R_{\ast}$) and for the star effective temperature ($T_{\rm eff}$) we used the formulas in 
\citep{2012ApJ...756...92C} for POP II stars and POP III stars.
We calculated separately the emission from POP II and POP III stars assuming that the POP III stellar population 
evolution could be described using the error function.
The error function is given by:

\be
fp(z')=\frac{1}{2}\left[ 1+erf\left(\frac{z-zt}{\sigma_p}\right)  \right],
\ee
\\
where we imposed that the POP III population ended at $z\, =\, 6$, that POP III stars are the dominant population 
for $z_{\rm t}\, \geq\, 10$ and that the POP III transition width is $\sigma_{\rm P}\,=\, 0.5$. A discussion for the 
choice of these values can be found at \citep{2013arXiv1301.6974F}.  

The observed stellar luminosity density is the sum of the luminosity density of POP II stars ($l^{\rm POPII}$) 
and of POP III stars ($l^{\rm POPIII}$) given respectively by:

\be
\label{eq:L_POPII}
l^{\rm POP_{II}}(\nu,z)=\int_z^{\rm z_{\rm max}} \int_{\rm M^{\rm min}_{\ast}}^{\rm M^{\rm cut(z,z')}_{\rm \ast}} K(z')\, M_{\rm \ast}^{-2.35}  L^{\star}_{\rm \nu}\, dM_{\rm \ast}'\, \frac{dt}{dz'}\, dz',
\ee
\\
and

\be
\label{eq:L_POPIII}
l^{\rm POP_{III}}(\nu,z)= \int_z^{\rm z_{max}} \int_{\rm M^{\rm min}_{\ast}}^{\rm M^{\rm cut(z')}_{\ast}} K(z')\, M_{\ast}^{-1} \left(1+\frac{M_{\ast}}{M^c_{\ast}}\right) ^{-1.35} L^{\ast}_{\rm \nu}\, dM_{\ast}\, \frac{dt}{dz'}\, dz',
\ee
\\
We integrated in stellar mass using a \citep{1955ApJ...121..161S} IMF (Initial Mass Function) with a 
mass range from 0.1 to 100 ${\rm M_{\sun}}$ for POP II stars and a \citep{1998MNRAS.301..569L} IMF with a mass 
range from 0.1 to 500 ${\rm M_{\sun}}$, and $M^c_{\ast}={\rm 250\, M_{\sun}}$ for POP III stars. 
In Equations \ref{eq:L_POPII} and \ref{eq:L_POPIII}, $M^{\rm cut}(z,z')$, corresponds to the maximum stellar 
mass of a star created at redshift $z'$ which is still alive at z and $K(z')$ is the normalization of the mass 
function in units of ${\rm [Mpc^{-3}\, s^{-1}]}$ so that the total stellar mass coincides with the value that can be 
obtain with the star formation rate density $\Psi$ in units of ${\rm M_{\odot}\, Mpc^{-3}\, s^{-1}}$.
For POP II and POP III stars $K(z')$ is given respectively by:

\ba
K(z')&=& \frac{\Psi(z')\, fp(z')}{ \int_{\rm M^{\rm min}_{\ast}}^{\rm M^{\rm max}_{\ast}} M_{\ast}^{-2.35}\, M_{\ast}\, dM_{\ast}'}
\ea
\\
and

\ba
K(z')&=& \frac{\Psi(z')[1-fp(z')]}{ \int_{\rm M^{\rm min}_{\ast}}^{\rm M^{\rm max}_{\ast}} M_{\ast}^{-1} \left(1+\frac{M_{\ast}}{M^c_{\ast}}\right) ^{-1.35}  M_{\ast}\, dM_{\ast}'}
\ea
\\
The stellar emission contamination to the observed frequency $\nu_o$ is given by:

\be
I_{\nu_o}=\int^{\rm z_{\rm max}}_{\rm z=0}\ y(z')\frac{l^{\rm POP_{\rm II}}(\nu,z')+l^{\rm POP_{\rm III}}(\nu,z')}{4\pi(1+z')^2} dz',
\ee
\\
where $\nu$ is the emitted frequency which is related to the observed frequency by $\nu\,=\, \nu_o\,(1\, +\, z)$.
The resulting intensity in the 200 GHz to 300 GHz frequency range is $I_{\rm \nu}\approx 2.5\, \times\, 10^{\rm -4}\, {\rm Jy\,  sr^{-1}}$.  

\subsubsection{Dust emission}
\label{Dust_emission}

UV emission from stars between $\rm 13.6\, eV$ and $\rm 6\, eV$ is absorbed by the dust in the galaxy and is re-emitted as continuum 
infrared radiation.
The dust spectral emission is the result of emission by dust particles with different sizes emitting at a temperature 
proportionally to the particle size. Each particle emits approximately as a black body and so the overall emission spectra
is well described by a black body spectra at temperature Td, modified by the emissivity function 
$\epsilon_{\nu}\, \propto\, \nu^{\beta}$ in order to account for the different dust temperatures.
Using the data from the Herschel-Astrophysical Terahertz Large Area Survey and from the Sloan Digital Sky Survey 
\cite{2010A&A...518L...9A} estimated that $\beta\, \approx\, 1.5$ and that the dust temperature as a function of the 
Infrared (IR) luminosity can be approximated by:

\be
T_d\, =\, T_0\, +\alpha\, {\rm log}\,(L_{\rm FIR}\, /\, {\rm L}_{\odot}),
\ee
\\
with $T_0\, =\, -20.5\, {\rm K}$ and $\alpha\, =\, 4.4$. The data set used for this fit contains galaxies with 
$L_{\rm IR}\, >\, 10^8\, {\rm L}_{\odot}$. For lower luminosity 
galaxies we assumed that $T_{\rm d}\, =\, T_{\rm d}(L_{\rm IR}\, =\, 10^8\, {\rm L}_{\odot})\, \approx\, 15\, {\rm K}$. 
We set the amplitude of the dust  emission ($A_{dust}$) by imposing that:

\be
A_{dust}=\frac{L_{IR}(1-1000\mu m)}{\int^{1000\mu m} _{1 \mu m}\,  \frac{\nu_e^{3+\beta}}{e^{(h \nu_e / k T_{\rm d})}\,-\,1} },
\ee
\\
where $L_{\rm IR}$ and $L_{\rm FIR}$ can be obtained from the galaxy mass using Equations \ref{eq:L_IR_SFR}, 
\ref{eq:LIR_LFIR} and \ref{eq:SFR_param}.
The observed intensity originated from dust emission will therefore be:

\be
I_{\nu}=\int_{\rm z_{\rm min}}^{\rm z_{\rm max}} \int_{\rm M_{\rm min}} ^{\rm M_{\rm max}} \frac{dn}{dm}\, \frac{A_{\rm dust}\, \frac{\nu_e^{3+\beta}}{e^{(h\nu_e/k T_{\rm d})}-1}}{4\, \pi\, D_{\rm L}^2} D_{\rm A}^2\,  y(\nu_e)\,  dM\, dz
\ee
\\
where $\nu_e\, =\, \nu\,  (1+z)$.
The resulting intensity is of the order of $2-3\, \times\, 10^{5}\, {\rm Jy\, sr^{-1}}$. This value is compatible with observations of the infrared background.
The contamination from dust emission will be orders of magnitude above the CII intensity in the 
200 - 300 GHz observing frequency range.  The continuum dust emission 
can however be efficiently subtracted from observations taking advantage of the smooth evolution of this foregroung compared to the CII intensity fluctuations \cite{2015arXiv150406530Y}.

\subsubsection{ Free-free, free-bound and two photon emission}
\label{sec:Free_emission}

Free electrons scatter off ions without being captured giving origin to free-free continuum emission.
If during this interaction the electrons are captured by the ions then there is emission of free-bound 
radiation. In these same ionized 
regions there occur hydrogen recombinations which originate lyman alpha photons or two photon emission.
Following the approach of \citep{2006ApJ...646..703F}, the free-free and free-bound and two photon continuum luminosities 
can be obtained using:

\be
L_{\nu}(M,z)=V_{sphere}(M,z)\, \varepsilon_{\nu}
\ee
\\
where $V_{\rm sphere}$ is the volume of the Str$\ddot{o}$mgren sphere (the emitting volume) which can be 
roughly estimated using the ratio between the number of ionizing photons emitted by a halo and the number density of 
recombinations in the ionized volume:

\be
V_{sphere}(M,z)=\frac{Q_{\rm ion}\, \psi(M,z)\, (1-f_{\rm esc})}{n_{\rm e}\, n_{\rm p}\, \alpha_{\rm B}},	
\ee
\\
where $\varepsilon_{\nu}$ is the total volume emissivity of free-free, free-bound emission and two photon emission, $Q_{\rm ion}$ is the 
average number of ionizing photons per solar mass in star formation, $n_{\rm e}$ is the number density of free electrons, 
$n_{\rm p}$ is the number density of protons (ionized atoms) and $\alpha_{\rm B}$ is the case B recombination coefficient 
(taken from \cite{2006PhR...433..181F}) given by:

\be
\alpha_B \approx 2.6 \times 10^{-13}\, (T_{\rm K}/10^4{\rm K})^{-0.7}\, (1+z)^3\, {\rm cm^3\, s^{-1}},
\ee
\\
and $f_{\rm esc}$ is the ionizing photon escape fraction. In our calculations we used the redshift and halo mass 
dependent escape fraction of ionizing radiation from \cite{2010ApJ...710.1239R} and we used 
$Q_{ion}\, =\, 5.38\, \times\, 10^{60}\, {\rm M_{\odot}^{-1}}$ appropriate for POP II stars \citep{2012arXiv1205.1493S}. 
\\
The volume emissivity from free-free and free-bound emission estimated by \citep{2003adu..book.....D} is given by:

\be
\varepsilon_{\nu}^{\rm free}=4\pi\, n_e\, n_p\, \gamma_c\, \frac{e^{-h\nu/kT_{\rm K}}}{T_{\rm K}^{1/2}}\, \rm J {\rm cm^{-3}\, s^{-1}\, Hz^{-1}}, 
\ee
\\
where $\gamma_c$ is the continuum emission coefficient for free-free and free-bound emission given in SI units by:

\be
\gamma_c=5.44\times 10^{-46}\left[\bar{g}_{ff}+\Sigma_{n=n^{\prime}}^{\infty}\, \frac{x_n e^{x_n}}{n}\, g_{fb}(n)\right],
\ee
\\
where $x_n\, =\, Ry\,/\,(k_B\, T_{\rm K}\, n^2)$ (n is the level to which the electron recombines to and 
$Ry\, =\, 13.6\, {\rm eV}$ is the Rydberg unit of energy), $\bar{g}_{ff}\, \approx\, 1.1-1.2$ and $g_{fb}(n)\, \approx\, 1.05-1.09$ are the 
thermally average Gaunt factors for free-free and free-bound emission (values from \cite{1961ApJS....6..167K}). The initial 
level $n^{\prime}$ is determined by the emitted photon frequency and satisfies the condition 
$c\, R_{\infty}\, /\, n^{\prime 2}\, <\, \nu\, <\, c\, R_{\infty}/(n^{\prime}-1)^2$ where $R_{\infty}\, =\, 1.1\, \times\, 10^7\, \rm m^{-1}$ is the Rydberg constant. 
We then obtained the free-free plus free bound luminosity formula given by:
\ba
L_{\nu}^{\rm free}(M,z)= 3.68\times 10^{16}\, \alpha_B^{-1}\, ({\rm T_K},z)\left[\frac{Q_{ion}}{5.38\times10^{60}}\right]\\ \nonumber
\left[1.15\, +\, \Sigma_{n=n^{\prime}}^{\infty}\, \frac{x_n\, e^{x_n}}{n}\, 1.07\right]\, \frac{e^{-h\nu/k_B T_{\rm K}}}{T_{\rm K}^{1/2}}\, \psi(M,z).
\ea
\\

The two photon luminosity is given by:

\be
L_{\nu}^{2\gamma}(M)\,=\, \frac{2h\nu}{\nu_{Ly\alpha}}\, (1\, -\, f_{Ly\alpha})\, P(\nu/\nu_{Ly\alpha})\,Q_{\rm ion}\,\psi(M,z),
\ee
\\
where P(y)dy is the normalized probability that in a two photon decay one of them is the range $dy\, =\, d\nu/\nu_{Ly\alpha}$ 
and $1\, -\, f_{Ly\alpha}\, \approx\, 1/3$ is the probability of 2 photon emission during an hydrogen n = 2 $\rightarrow$ 1 transition. 
The probability of two photon decay was fitted by \cite{2006ApJ...646..703F} using Table 4 of \citet{1970ApJ...160..939B} as:

\be
P(y)\, =\, 1.307\, -\, 2.627\, (y\, -\, 0.5)^2\, +\, 2.563\,(y-0.5)^4\, -\, 51.69\, (y-0.5)^6.
\ee
\\

The intensity from nebula continuum emission is given by:

\be
I_{\rm cont}(\nu)\, =\,  \int_{z=0} ^{z_{\rm max}}\, dz\, \frac{L_{\rm cont}(\nu_e,z)}{4\pi\, D_{\rm L}^2}\, y\, D_{\rm A}^2 ,
\label{eq:I_ave}
\ee
\\
where $L_{\rm cont}\, =\, L_{\nu}^{\rm 2\gamma}\,  +\,  L_{\nu}^{\rm free} $.

Assuming a gas temperature of $(1-4)\, \times\, 10^4\, {\rm K}$ we estimated the intensity of nebula continuum emission in the frequency range 
200 - 300 GHz to be of $I_{\rm cont}(\nu)=I_{\rm free}+I_{\rm 2\gamma}\approx\, (0.9-1.3)\, \times\, 10^{1}\,{\rm Jy\, sr^{-1}} +  (1.8-7.2)\, \times\, 10^{-13}\, {\rm Jy\, sr^{-1}} =(0.9-1.3)\, \times 10^{1}  {\rm Jy\, sr^{-1}}$. 
Therefore the nebula continuum emission is dominated by free-free emission and is below the expected CII signal. This continuum can easily be removed 
from observations taking advantage of the smooth evolution of this foreground with frequency such as in the case of dust continuum emission.

\end{document}